\documentclass[aps,prd,onecolumn,groupedaddress,showpacs,nofootinbib,amssymb]{revtex4-2}
\usepackage[dvips]{graphicx}
\usepackage{amssymb}
\usepackage{amsmath}
\usepackage{graphicx,,color}
\usepackage{amsfonts}
\usepackage{bm}
\usepackage{cancel}
\usepackage{comment}

\newcommand\be{\begin{equation}}
\newcommand\ee{\end{equation}}

\allowdisplaybreaks[4]

\begin{document}

\tolerance=5000

\title{Quantitative Predictions for $f(R)$ Gravity Primordial Gravitational Waves}
\author{S.D.~Odintsov,$^{1,2}$}
\email{odintsov@ice.cat}
\author{V.K.~Oikonomou,$^{3,4}$}
\email{v.k.oikonomou1979@gmail.com,voikonomou@auth.gr}
\author{F.P. Fronimos,$^{3}$}
\email{fotisfronimos@gmail.com,ffronimo@physics.auth.gr}
\affiliation{$^{1)}$ ICREA, Passeig Luis Companys, 23, 08010 Barcelona, Spain\\
$^{2)}$ Institute of Space Sciences (IEEC-CSIC) C. Can Magrans
s/n,
08193 Barcelona, Spain\\
$^{3)}$ Department of Physics, Aristotle University of
Thessaloniki, Thessaloniki 54124,
Greece\\
$^{4)}$ Laboratory for Theoretical Cosmology, Tomsk State
University of Control Systems and Radioelectronics, 634050 Tomsk,
Russia (TUSUR)\\
}

 \tolerance=5000

\begin{abstract}
In this work we shall develop a quantitative approach for
extracting predictions on the primordial gravitational waves
energy spectrum for $f(R)$ gravity. We shall consider two distinct
models which yield different phenomenology, one pure $f(R)$
gravity model and one Chern-Simons corrected potential-less
$k$-essence $f(R)$ gravity model in the presence of radiation and
non-relativistic perfect matter fluids. The two $f(R)$ gravity
models were carefully chosen in order for them to describe in a
unified way inflation and the dark energy era, in both cases
viable and compatible with the latest Planck data. Also both
models mimic the $\Lambda$-Cold-Dark-Matter model and specifically
the pure $f(R)$ model only at late times, but the Chern-Simons
$k$-essence model during the whole evolution of the model up to
the radiation domination era. In addition they guarantee a smooth
transition from the inflationary era to the radiation, matter
domination and subsequently to the dark energy era. Using a WKB
approach introduced in the relevant literature by Nishizawa, we
derive formulas depending on the redshift that yield the modified
gravity effect, quantified by a multiplicative factor, a
``damping'' in front of the General Relativistic waveform. In
order to calculate the effect of the modified gravity, which is
the ``damping'' factor, we solve numerically the Friedmann
equations using appropriate initial conditions and by introducing
specific statefinder quantities. As we show, the pure $f(R)$
gravity gravitational wave energy spectrum is slightly enhanced,
but it remains well below the sensitivity curves of future
gravitational waves experiments. In contrast, the Chern-Simons
$k$-essence $f(R)$ gravity model gravitational wave energy
spectrum is significantly enhanced and two signals are predicted
which can be verified by future gravitational wave experiments. We
discuss in detail our findings and the future perspective of
modified gravity theories in view of the upcoming second and third
generation experiments on primordial gravitational waves.
\end{abstract}


\maketitle

\section{Introduction}

Inflation \cite{inflation1,inflation2,inflation3,inflation4} is to
date the most promising candidate theory for describing the
post-Planckian epoch of our Universe. The inflationary era is
speculated to have occurred after the quantum gravity era of our
Universe, and from the beginning of the inflationary era the
Universe is assumed to be four dimensional and to be described by
classical physics. The inflationary scenario is appealing since
most inflationary theories generate a nearly scale invariant power
spectrum of the primordial scalar perturbations, and it solves the
most unappealing issues of standard Big Bang cosmology, such as
the horizon and flatness problems. However, to date, the direct
verification that inflation has ever occurred has not been
accomplished yet. This direct verification of the inflationary era
can be given only via the detection of the $B$-modes (curl modes)
of inflation in the Cosmic Microwave Background (CMB) radiation
temperature and polarization anisotropies
\cite{Kamionkowski:2015yta}. The CMB probes modes with wavenumbers
$k<0.62$$\,$Mpc$^{-1}$, and for modes with wavenumbers larger than
this cutoff it is impossible for the CMB to probe them, because
for $f>0.15\times 10^{-15}$Hz the scalar perturbations are highly
non-linear, thus CMB probes scales with wavelength $\lambda$ from
$10$$\,$Mpc to $10^4$$\,$Mpc. $B$-mode polarization can mainly
arise from two sources in the Universe, firstly from the
primordial tensor perturbations, the inflationary tensor modes,
which occur at the low-multipoles ($\ell \leq 10$) or equivalently
large angular scales of the CMB radiation, or from the
gravitational lensing conversion of the $E$-mode polarization
modes into the (curl) $B$-modes, and this physical process occurs
at late times and on small angular scales
\cite{Denissenya:2018mqs}. Thus, the goal of the next generation
of CMB experiments has become the detection of CMB polarization
patterns induced by inflationary tensor modes with wavelengths
from $10$$\,$Mpc to $10^4$$\,$Mpc. It is conceivable that the CMB
probes modes entered the horizon well after the Big Bang
Nucleosynthesis (BBN) epoch which occurred for temperatures $T\sim
1$MeV with wavenumber $k_{BBN}=6.5\times 10^3$$\,$Mpc$^{-1}$. Thus
in some sense, the CMB offers insights for the physics that
occurred during the last stages of the matter domination era,
beyond the matter-radiation equality $k_{eq}\sim
0.05\,h^2$$\,$Mpc$^{-1}$, leaving the physics of the early matter
and radiation-reheating era untouched. The main reason for this is
the fact that scalar perturbations become non-linear for
wavelengths below $10$$\,$Mpc. The modes with wavelength well
below $10$$\,$Mpc have entered the horizon well before the
matter-radiation equality era, and these correspond to primordial
tensor modes that probe the radiation and reheating era
\cite{Kamionkowski:2015yta,Denissenya:2018mqs,Turner:1993vb,Boyle:2005se,Schutz:2010xm,Sathyaprakash:2009xs,Caprini:2018mtu,
Arutyunov:2016kve,Kuroyanagi:2008ye,Clarke:2020bil,Kuroyanagi:2014nba,Nakayama:2009ce,Smith:2005mm,Giovannini:2008tm,
Liu:2015psa,Zhao:2013bba,Vagnozzi:2020gtf,Watanabe:2006qe,Kamionkowski:1993fg,Giare:2020vss,Kuroyanagi:2020sfw,Zhao:2006mm,
Nishizawa:2017nef,Arai:2017hxj,Bellini:2014fua,Nunes:2018zot,DAgostino:2019hvh,Mitra:2020vzq,Kuroyanagi:2011fy,Campeti:2020xwn,
Nishizawa:2014zra,Zhao:2006eb,Cheng:2021nyo,Nishizawa:2011eq,Chongchitnan:2006pe,Lasky:2015lej,Guzzetti:2016mkm,Ben-Dayan:2019gll,
Nakayama:2008wy,Capozziello:2017vdi,Capozziello:2008fn,Capozziello:2008rq,Cai:2021uup}.
These inflationary modes have reentered the Hubble horizon after
the end of the inflationary era first and this occurred during the
reheating and radiation era. For these eras we basically know
nothing and the only way to probe these eras is via the primordial
gravitational waves. These inflationary waves form a stochastic
background of inflationary tensor modes and carry important
information about the conditions and state of our Universe during
and after the inflationary era. The stochastic gravitational
background is also strongly affected by the evolution of the
Universe and is also affected by the matter content of our
Universe after inflation. These effects imprinted on the
stochastic gravitational wave background radiation offer
incredible insights for the reheating and radiation domination
eras and specifically between the end of inflation and the
electroweak phase transition. The appealing feature of the
stochastic gravitational wave background is that the primordial
gravity waves have tiny amplitude and their interaction with
matter is significantly small, which makes them obey linear
evolution equations. Hence, their evolution is easier to predict,
in contrast to density perturbations which are non-linear for
modes larger than $10$$\,$Mpc and grow at superhorizon scales. The
primordial tensor modes are mainly affected by the first horizon
crossing during inflation, also by the horizon reentry after
inflation, and by the evolution itself of the Universe and of the
matter content of the Universe. Indirectly, the effects of a
modified gravity governing the evolution are imprinted also on
primordial gravitational waves, since a modified gravity may
directly affect the inflationary and post-inflationary era. Thus
via the stochastic primordial gravitational waves background,
which are nothing else but the superadiabatic amplified zero point
fluctuations of the gravitational field, the physics of the
post-inflationary early Universe might be uniquely probed and
studied. If we will be able to probe these primordial tensor
modes, we will be able to lay hands on cosmological eras which are
currently unknown for us. Also let us note that the lowest
frequency mode observable today corresponds to modes that have
entered the Hubble horizon at present day, and these modes can be
probed by the CMB. Basically, it is useful to understand the
spectrum of frequencies existing in our Universe today. As we
already mentioned, the lowest frequency mode probed by the CMB is
the mode that enters the horizon today, and the maximum frequency
of the primordial gravitational waves spectrum is $f_{max}\sim
1$GHz \cite{Giovannini:2008tm}. The physics in the range
$f_{BBN}<f<f_{max}$ is pretty much unknown thus the primordial
gravitational waves actually reveal information related to several
features of this physically unknown era, not up to $f_{max}$ but
for frequencies up to several Hz. The maximum frequency is hard to
probe, but future experiments related to radiation and reheating
era modes, will probe frequencies up to several hundreds of Hz.

The future experiments, second and third generation experiments,
that will be able to probe modes that entered the Hubble horizon
at temperatures much higher than the BBN one, are: ground-based
interferometers, like the Einstein Telescope probing Hz-KHz
frequencies \cite{Hild:2010id}, space-born interferometers like
the LISA laser interferometer space antenna
\cite{Baker:2019nia,Smith:2019wny}, future space missions like the
BBO (Big Bang Observer) \cite{Crowder:2005nr,Smith:2016jqs}, the
DECIGO \cite{Seto:2001qf,Kawamura:2020pcg} and also the SKA
(Square Kilometer Array) which will receive data from pulsar
timing arrays at frequencies $10^{-8}$Hz \cite{Bull:2018lat}. Also
currently and in the future the NANOGrav
\cite{Arzoumanian:2020vkk,Pol:2020igl} provides data on pulsar
timing arrays too.

The future experiments like the BBO, DECIGO and LISA and the
Einstein Telescope will probe frequencies that peak near the
frequency range $f\sim \mathcal{O}(10^{-4}-1)$Hz, which basically
corresponds to tensor modes with $k\sim
\mathcal{O}(10^{10}-10^{15})$$\,$Mpc$^{-1}$. These modes
correspond to modes that have reentered the Hubble horizon well
before the matter-radiation equality which in turn corresponds to
modes with $k_{eq}\sim 0.05\,h^2$$\,$Mpc$^{-1}$. Also the
aforementioned future missions and experiments probe modes well
beyond the tensor modes that have reentered the Hubble horizon
during the BBN era, at $T\sim 1$MeV with wavenumber
$k_{BBN}=6.5\times 10^3$$\,$Mpc$^{-1}$. The maximum frequency of
the primordial gravitational waves spectrum is $f_{max}\sim 1$GHz
\cite{Giovannini:2008tm}. We need to note that the gravitational
waves below $f\sim 10^{-15}$Hz are affected by a second order term
of the scalar primordial perturbations, via the anisotropic stress
term, and these effects cause a damping of the energy spectrum,
but these frequencies are not of interest for future experiments
which probe much higher frequencies.

The standard theory that is most frequently used for inflationary
theories is the scalar field theory approach
\cite{inflation1,inflation2,inflation3,inflation4}. However the
scalar field description offers a very restricted description of
inflation, which can also be connected with theoretical
shortcomings like the Swampland issue
\cite{Vafa:2005ui,Ooguri:2006in,Palti:2020qlc,Odintsov:2020zkl}.
In addition, in the context of scalar field inflation it is
usually quite difficult to describe in a unified way inflation and
the dark energy era. In contrast, in the context of modified
gravity in its various forms
\cite{reviews1,reviews2,reviews3,reviews4,reviews5,reviews6}, the
unified description of inflation and the dark energy era can be
accomplished, see for example the pioneer work in the context of
$f(R)$ gravity \cite{Nojiri:2003ft} and for later developments see
\cite{Nojiri:2007as,Nojiri:2007cq,Cognola:2007zu,Nojiri:2006gh,Appleby:2007vb,Elizalde:2010ts,Odintsov:2020nwm,Oikonomou:2020qah,Oikonomou:2020oex}.
With regard to primordial gravitational waves, it seems that
scalar field theories remain beyond the reach of future
experiments, since the scalar field theory primordial
gravitational wave energy spectrum lies well below the sensitivity
curves of most of the future experiments \cite{Breitbach:2018ddu}.
In fact, the source of this problem seems to be the red-tilted
tensor spectral index predicted by the scalar field theories. To
date it is becoming a fact that a blue-tilted tensor spectral
index will generate a gravitational wave energy spectrum that may
be detected in the future second and third generation experiments
on gravitational waves \cite{Kuroyanagi:2020sfw,Vagnozzi:2020gtf}.
In scalar field theory inflation, a blue-tilted tensor spectral
index cannot be generated by a non-tachyonic theory
\cite{Zhao:2013bba}, a factor that narrows the possibility of
detection of scalar field theory inflation if indeed it describes
inflation, if the latter ever occurred. Thus if inflation occurred
primordially, and a signal of stochastic tensor perturbation is
observed in the future experiments, single scalar field
inflationary theory will be overruled from being a viable
description of inflation. To this end modified gravity combined or
not with scalar fields may provide a viable candidate theory for
describing inflation. Motivated by this aspect, in this work we
will study primordial gravitational waves in the context of $f(R)$
gravity. After providing some essential information about the
General Relativistic (GR) waveform of the stochastic gravitational
waves, we shall employ a WKB approach introduced by Nishizawa
\cite{Nishizawa:2017nef} in order to provide a quantifiable way to
measure the effects of modified gravity on the primordial
gravitational wave waveform. The resulting waveform contains the
GR waveform multiplied by a ``damping'' factor, which is solely
affected by the modified gravity controlling the evolution. This
``damping'' factor might actually cause an overall damping of the
gravitational wave energy spectrum or might enhance it, depending
on the underlying modified gravity controlling the evolution. The
only way to calculate this ``damping'' factor is by solving the
Friedmann equation numerically for the physically relevant
redshift range, which extends from present day up to the radiation
domination era. We shall introduce suitable statefinder
quantities, dark energy based, and we shall transform the
Friedmann equations appropriately in order to express them in
terms of these statefinders and in terms of the redshift. Our aim
is to study two distinct $f(R)$-gravity related models, firstly a
pure $f(R)$ gravity model and secondly a Chern-Simons
potential-less $k$-essence $f(R)$ gravity, both in the presence of
dark matter and radiation perfect fluids. For the pure $f(R)$
gravity case, the contribution of the modified gravity on the
`damping'' term basically is absent beyond redshift $z\sim 1000$,
and also for redshifts for which the total equation of state (EoS)
parameter reaches the radiation domination value $\omega_{tot}\sim
1/3$, the ``damping'' term is zero. Thus we provide a calculation
of the ``damping'' factor and the result is that the actual pure
$f(R)$ gravitational wave spectrum is slightly enhanced compared
to the GR one, but still remains well below the sensitivity curves
of the future experiments. The possibility of having enhanced
gravitational wave spectrum in the context of $f(R)$ gravity has
also been stressed in Refs.
\cite{Capozziello:2017vdi,Capozziello:2008fn,Capozziello:2008rq}.
We perform the same analysis for the Chern-Simons $k$-essence
$f(R)$ gravity, and as we will show, the primordial gravitational
wave energy spectrum is significantly enhanced, and can be
detectable by most of the future experiments. More importantly, we
show that due to the Chern-Simons term, the prediction for the
theory is actually not one, but two distinct signals peaking at
the same frequency range. Also we demonstrate that the reheating
temperature significantly affects the gravitational wave energy
spectrum. Thus by using two appropriate toy models, we show in a
quantitative way what may occur in future experiments. For both
the models we considered, we used models that may describe the
dark energy and the inflationary era within the same theoretical
framework, so the models we used are severely constrained
phenomenologically. Moreover as a side remark, the models we used
were actually able to also describe the intermediate eras in
between the inflationary and the dark energy era, that is, the
matter and radiation domination eras, and a smooth transition
between all cosmological eras. Also one of the models behaves
exactly as the $\Lambda$-Cold-Dark-Matter ($\Lambda$CDM) model. We
also discuss the future observations outcomes, in view of our
findings and we theorize what a detection or a non-detection of a
signal would mean for inflationary physics.

This work is organized as follows: In section II we present the
essential features of the primordial gravitational waves in the
context of GR and we also present how to apply the WKB method in
order to extract the modified gravity effects on the primordial
gravitational waves waveform. We consider two cases, the pure
$f(R)$ gravity case and a Chern-Simons corrected $k$-essence
$f(R)$ gravity, and we provide explicit formulas in both cases for
the ``damping'' factor and the primordial gravitational wave
energy spectrum. In the same section we describe in brief our
strategy for obtaining the ``damping'' factor and we discuss our
aim to present two models that can describe inflation and dark
energy in a unified way. In section III we shall study two
distinct models of $f(R)$ gravity with respect to the primordial
gravitational wave energy spectrum, firstly a pure $f(R)$ gravity
model and secondly a Chern-Simons $k$-essence $f(R)$ gravity
model, both in the presence of dark matter and radiation perfect
fluids. In both cases the models were appropriately chosen in
order for them to provide a unified description of inflation and
the dark energy eras. We provide for both the models detailed
description for the dark energy era, and beyond, by solving
numerically the Friedmann equation using appropriate initial
conditions, and we also discuss the inflationary aspects of the
models. Accordingly we calculate for both models the ``damping''
factor and the overall effect of modified gravity on the
primordial gravitational wave energy spectrum. We also provide
detailed predictions for the produced gravitational wave energy
spectrum and we compare our findings with the projected
sensitivities of the most future experiments, for a wide range of
frequencies. We also discuss in some detail what implications for
inflation would have the absence or presence of a signal in future
experiments. Finally, the concluding remarks along with a critical
discussion on our findings are presented in the conclusions
section.

\section{Primordial Gravitational Waves in Einstein-Hilbert and $f(R)$ Gravity}

\subsection{The Einstein-Hilbert Gravity Case}

Before getting to the core of this work, it is vital to present
some general features of primordial gravitational waves in the
context of standard general relativity (GR) and in the context of
$f(R)$ gravity. We shall adopt the notation and conventions of
Refs.
\cite{Boyle:2005se,Nishizawa:2017nef,Arai:2017hxj,Nunes:2018zot,Liu:2015psa,Zhao:2013bba}
and references therein. The material that will be presented is not
new but it will help the article to be self-contained.

Let us start with the GR description of inflationary gravitational
waves. For our analysis we shall assume a spatially flat
Friedmann-Robertson-Walker (FRW) metric with line element,
\begin{equation}
\label{metric} \centering {\rm d}s^2=-{\rm
d}t^2+a(t)^2\sum_{i=1}^{3}{({\rm d} x^{i})^2}\, ,
\end{equation}
where $a(t)$ denotes the scale factor and $t$ denotes the cosmic
time. For inflationary gravitational waves considerations, it is
usually more convenient to use the conformal time $\tau$, in which
case the line element of the FRW spacetime reads,
\begin{equation}
  {\rm d}s^{2}=a^{2}[-{\rm d}\tau^{2}+(\delta_{ij}+h_{ij})
  {\rm d}x^{i}{\rm d}x^{j}],
\end{equation}
and $x^{i}$ denotes the comoving spatial coordinates, while
$h_{ij}$ stands for the gauge-invariant metric tensor
perturbation. The metric tensor perturbation $h_{ij}$ is symmetric
($h_{ij}\!=\!h_{ji}$), and satisfies the traceless $h_{ii}\!=\!0$,
and transverse $\partial^j h_{ij}\!=\!0$ conditions. Treating the
tensor perturbation $h_{ij}(\tau,{\bf x})$ as a quantum field
embedded in an unperturbed FRW spacetime $g_{\mu\nu}
=diag\{-a^{2},a^{2},a^{2},a^{2}\}$, and by keeping order two terms
in $h_{ij}$ in the Lagrangian of the gravitational field, the
tensor perturbations quadratic action is given by,
\begin{equation}
  \label{tensor_action}
  S=\int d\tau d{\bf x}\sqrt{-g}\left[
    \frac{-g^{\mu\nu}}{64\pi G}\partial_{\mu}h_{ij}
    \partial_{\nu}h_{ij}+\frac{1}{2}\Pi_{ij}h_{ij}\right].
\end{equation}
with $g^{\mu\nu}$ being the inverse of $g_{\mu\nu}$ and $g$
denoting its determinant. The anisotropic stress $\Pi_{\mu \nu}$,
with its tensor part being,
\begin{equation}
  \Pi^{i}_{j}=T^{i}_{j}-p\delta^{i}_{j}
\end{equation}
satisfies $\Pi_{ii}=0$ and the transverse condition
$\partial^{i}\Pi_{ij}=0$, and its coupling to the tensor
perturbation $h_{ij}$ acts like an external source in the
gravitational quantum action (\ref{tensor_action}). Upon variation
of the action (\ref{tensor_action}) with respect to the tensor
perturbation $h_{ij}$, we get the following equation of motion,
\begin{equation}
  \label{h_eq}
  h_{ij}''+2\frac{a'(\tau)}{a(\tau)}h_{ij}'-{\bf \nabla}^{2}h_{ij}
  =16\pi G a^{2}(\tau)\Pi_{ij}(\tau,{\bf x}),
\end{equation}
where the prime denotes differentiation with respect to the
conformal time $\tau$. Performing a Fourier transform in the
equation of motion (\ref{h_eq}), we obtain,
\begin{subequations}
  \label{fourier_expand}
  \begin{eqnarray}
    h_{ij}^{}(\tau,{\bf x})\!\!&=\!&\!\!\sum_{r}
    \sqrt{16\pi G}\!\!\int\!\!\!
    \frac{d{\bf k}}{(2\pi)^{3/2}}\epsilon_{ij}^{r}({\bf k})
    h_{{\bf k}}^{r}(\tau){\rm e}^{i{\bf k}{\bf x}},\qquad\quad \\
    \Pi_{ij}^{}(\tau,{\bf x})\!\!&=\!&\!\!\sum_{r}
    \sqrt{16\pi G}\!\!\int\!\!\!
    \frac{d{\bf k}}{(2\pi)^{3/2}}\epsilon_{ij}^{r}({\bf k})
    \Pi_{{\bf k}}^{r}(\tau){\rm e}^{i{\bf k}{\bf x}},\qquad\quad
  \end{eqnarray}
\end{subequations}
where $r=$(``$+$'' or ``$\times$'') indicates the polarization of
the gravitational tensor perturbation. We need to note that the
Fourier transform is essential in order to obtain the evolution
equation corresponding to each mode quantified by the wavenumber
$\vec{k}$. Also, Eq. (\ref{h_eq}) describes the evolution of the
metric tensor perturbation, which depends on space and time, but
for the analysis of the primordial gravitational waves, it is
required to obtain the behavior of each mode of specific frequency
and wavenumber $\vec{k}$. Each Fourier mode has only time
dependence, thus it's evolution can be split in several distinct
evolutionary eras, plus the subhorizon and superhorizon behavior
can easily be obtained.

Also the polarization tensors satisfy [$\epsilon_{ij}^{r} ({\bf
  k})=\epsilon_{ji}^{r}({\bf k})$], and also satisfy the traceless $\epsilon_{ii}^{r}
({\bf k})=0$, and the transverse conditions
$k_{i}\epsilon_{ij}^{r}({\bf k})=0$, as it is expected. By
choosing a circular-polarization basis in which
$\epsilon_{ij}^{r}({\bf k})=(\epsilon_{ij}^{r} (\textrm{-}{\bf
  k}))^{\ast}$, we can normalize the polarization basis in the
  following way,
\begin{equation}
  \label{basis_norm}
  \sum_{i,j}\epsilon_{ij}^{r}({\bf k})(\epsilon_{ij}^{s}
  ({\bf k}))^{\ast}=2\delta^{rs}.
\end{equation}
Combining Eq. (\ref{fourier_expand}) and (\ref{tensor_action}), we
obtain,
\begin{equation}
  \label{tensor_action_fourier}
  S\!=\!\!\sum_{r}\!\!\int\!\!d\tau d{\bf k}\frac{a^{2}}{2}\;\!\!
  \Big[h_{{\bf k}}^{r}{}'h_{\!\textrm{-}{\bf k}}^{r}\!{}'\!
  -\!k^{2}h_{{\bf k}}^{r}h_{\!\textrm{-}{\bf k}}^{r}\!
  +\!32\pi G a^{2}\Pi_{{\bf k}}^{r}h_{\!\textrm{-}{\bf k}}^{r}
  \Big]\, ,
\end{equation}
so essentially we obtain the original action of the tensor
perturbations (\ref{tensor_action}) expressed in terms of the
Fourier transformed tensor perturbations. This is essential for
the quantization of the resulting action (\ref{fourier_expand}),
in order to impose the isochronous commutation relations for each
distinct wavenumbers $\vec{k}$ and $\vec{k}'$. These relations are
much more simplified in Fourier space and can easily result to the
well-known quantum algebra for the bosonic creation and
annihilation operators. So we can proceed to the canonical
quantization of the above action, with $h_{{\bf k}}^{r}$ being the
canonical variable and its conjugate momentum being
\begin{equation}
  \label{def_pi}
  \pi_{{\bf k}}^{r}(\tau)=a^{2}(\tau)h_{\!\textrm{-}{\bf
  k}}^{r}{}'(\tau)\, ,
\end{equation}
so by promoting these variables to quantum operators
$\hat{h}_{{\bf k}}^{r}$ and $\hat{\pi}_{{\bf k}}^{r}$, we can
quantize the theory by imposing the equal-time commutation
relations,
\begin{subequations}
  \label{h_pi_commutators}
  \begin{eqnarray}
    \left[\hat{h}_{{\bf k}}^{r}(\tau),
      \hat{\pi}_{{\bf k}'}^{s}(\tau)\right]
    &=&i\delta^{rs}\delta^{(3)}({\bf k}-{\bf k}'), \\
    \left[\hat{h}_{{\bf k}}^{r}(\tau),
      \hat{h}_{{\bf k}'}^{s}(\tau)\right]
    &=&\left[\hat{\pi}_{{\bf k}}^{r}(\tau),
      \hat{\pi}_{{\bf k}'}^{s}(\tau)\right]=0.
  \end{eqnarray}
\end{subequations}
Note that the Fourier components of $\hat{h}_{ij}(\tau,{\bf x})$
satisfy $\hat{h}_{{\bf k}}^{r}=\hat{h}_{\!\textrm{-}{\bf
k}}^{r\dag}$, due to the fact that $\hat{h}_{ij}(\tau,{\bf x})$ is
Hermitian. Accordingly we can write these as follows,
\begin{equation}
  \label{h_from_a}
  \hat{h}_{{\bf k}}^{r}(\tau)=h_{k}^{}(\tau)\hat{a}_{{\bf k}}^{r}
  +h_{k}^{\ast}(\tau)\hat{a}_{\!\textrm{-}{\bf k}}^{r\dag},
\end{equation}
with the $\hat{a}_{{\bf k}}^{r\dag}$ being the creation operators
and  $\hat{a}_{{\bf k}}^{r}$ being the annihilation operators,
which satisfy the standard commutation relations,
\begin{subequations}
  \label{a_commutators}
  \begin{eqnarray}
    \Big[\hat{a}_{{\bf k}}^{r},\hat{a}_{{\bf k}'}^{s\dag}\Big]
    &=&\delta^{rs}\delta^{(3)}({\bf k}-{\bf k}'), \\
    \Big[\hat{a}_{{\bf k}}^{r},\hat{a}_{{\bf k}'}^{s}\Big]
    &=&\Big[\hat{a}_{{\bf k}}^{r\dag},\hat{a}_{{\bf
    k}'}^{s\dag}\Big]=0\, .
  \end{eqnarray}
\end{subequations}
Also the linearly-independent modes $h_{k}(\tau)$ and
$h_{k}^{\ast}(\tau)$ are solutions of the Fourier transformed
equation of motion,
\begin{equation}
  \label{h_eq_ft}
  h_{k}''+2\frac{a'(\tau)}{a(\tau)}h_{k}'+k^{2}h_{k}^{}=
    16\pi G a^{2}(\tau)\Pi_{k}^{}(\tau).
\end{equation}
From Eq.\ (\ref{h_from_a}) it is apparent that the modes
$h_{k}^{}(\tau)$ depend on the conformal time and on the
wavenumber $k=|{\bf k}|$, and not on the polarization and the
direction. The compatibility constraint between the two
commutation relations (\ref{h_pi_commutators}) and
(\ref{a_commutators}), is basically the following Wronskian
normalization condition,
\begin{equation}
  \label{Wronskian}
  h_{k}^{}(\tau)h_{k}^{\ast}{}'(\tau)-h_{k}^{\ast}(\tau)h_{k}'(\tau)
  =\frac{i}{a^{2}(\tau)}
\end{equation}
in the chronological past. The initial condition chosen in the
literature in the chronological past is the Bunch-Davies vacuum,
\begin{equation}
  \label{h_bc}
  h_{k}^{}(\tau)\to\frac{{\rm exp}(-ik\tau)}{a(\tau)\sqrt{2k}}
  \qquad({\rm as}\;\;\tau\to-\infty),
\end{equation}
which describes the modes $\vec{k}$ which are still in subhorizon
scales during inflation, and of course satisfies relation
(\ref{Wronskian}). For the quantitative study of the stochastic
primordial gravitational wave background, we shall use two
frequently used in the literature spectra, the tensor power
spectrum $\Delta_{h}^{2}(k,\tau)$ and the energy spectrum of
primordial gravitational waves $\Omega_{gw}^{}(k,\tau)$. We have,
\begin{equation}
  \langle0|\hat{h}_{ij}^{}(\tau,{\bf x})\hat{h}_{ij}^{}
  (\tau,{\bf x})|0\rangle\!=\!\!\!\int_{0}^{\infty}\!\!\!\!\!64\pi G
  \frac{k^{3}}{2\pi^{2}}\!\left|h_{k}^{}(\tau)\right|^{2}
  \!\frac{dk}{k},
\end{equation}
and therefore, the inflationary tensor power spectrum is,
\begin{equation}
  \label{def_tensor_power}
  \Delta_{h}^{2}(k,\tau)\equiv\frac{d\langle0|\hat{h}_{ij}^{2}
    |0\rangle}{d\,{\rm ln}\,k}=64\pi G\frac{k^{3}}{2\pi^{2}}
  \left|h_{k}^{}(\tau)\right|^{2}.
\end{equation}
 The present day energy spectrum of the stochastic primordial
 gravitational waves background $\Omega_{gw}^{}(k,\tau)$ is given
 by,
\begin{equation}
  \label{def_Omega_gw}
  \Omega_{gw}^{}(k,\tau)\equiv\frac{1}{\rho_{crit}^{}(\tau)}
  \frac{d\langle 0|\hat{\rho}_{gw}^{}(\tau)|0\rangle}{d\,{\rm ln}\,k}
\end{equation}
and it basically corresponds to the gravitational-wave energy
density ($\rho_{gw}^{}$) calculated per logarithmic wavenumber
interval, and the critical energy density is given by,
\begin{equation}
  \label{def_rho_crit}
  \rho_{crit}^{}(\tau)=\frac{3H^{2}(\tau)}{8\pi G}.
\end{equation}
In order to calculate $\Omega_{gw}^{}(k,\tau)$, we can assume that
the tensor perturbation $h_{ij}$ is a quantum field in the
unperturbed FRW background, and by using its action
(\ref{tensor_action}), we can calculate its stress-energy tensor,
which is,
\begin{equation}
  \label{T_alpha_beta}
  T_{\alpha\beta}=-2\frac{\delta L}{\delta\bar{g}^{\alpha\beta}}
  +\bar{g}_{\alpha\beta}L,
\end{equation}
where $L$ stands for the Lagrangian function in
(\ref{tensor_action}). Therefore, the gravitational wave energy
density in the absence of anisotropic stress couplings (this
simplification is justified for frequencies larger than the CMB
frequencies) is,
\begin{equation}
  \label{rho_gw}
  \rho_{gw}^{}=-T_{0}^{0}=\frac{1}{64\pi G}
  \frac{(h_{ij}')^{2}+(\vec{{\bf \nabla}}h_{ij})^{2}}{a^{2}},
\end{equation}
with its vacuum expectation value being,
\begin{equation}
  \label{rho_gw_expect}
  \langle0|\rho_{gw}^{}|0\rangle=\int_{0}^{\infty}\frac{k^{3}}
  {2\pi^{2}}\frac{\left|h_{k}'\right|^{2}
    +k^{2}\left|h_{k}^{}\right|^{2}}{a^{2}}\frac{dk}{k},
\end{equation}
and therefore the stochastic gravitational wave energy spectrum is
given by,
\begin{equation}
  \Omega_{gw}^{}(k,\tau)=\frac{8\pi G}{3H^{2}(\tau)}
  \frac{k^{3}}{2\pi^{2}}
  \frac{\left|h_{k}'(\tau)\right|^{2}
    +k^{2}\left|h_{k}^{}(\tau)\right|^{2}}{a^{2}(\tau)}.
\end{equation}
Furthermore, by using
$|h_{k}'(\tau)|^{2}=k^{2}|h_{k}^{}(\tau)|^{2}$, we may write the
gravitational wave energy spectrum at present day as,
\begin{equation}
  \label{spec_relations}
  \Omega_{gw}(k,\tau)=\frac{1}{12}\frac{k^{2}\Delta_{h}^{2}(k,\tau)}
  {H_0^{2}(\tau)}\, .
\end{equation}
where we assumed that the scale factor at present day is $a_0=1$
and $H_0$ stands for the present day Hubble rate, which is
$H_0\sim 67.3$ according to the latest Planck CMB based
constraints. There is a tension between the CMB-based value of
$H_0$ and the one indicated by the Cepheids, however in the
present paper we shall take into account only the CMB-based value
for the current Hubble rate, until several issues related to the
calibration of data for the Cepheids observations are consistently
and firmly resolved
\cite{Mortsell:2021nzg,Perivolaropoulos:2021bds}. Once the mode
with comoving wave number $k$ becomes superhorizon evolving, the
amplitude of the gravitational wave remains constant and when it
reenters the horizon, the amplitude is dumped. In the present work
we shall consider modes that became subhorizon at some point after
inflation, during the mysterious reheating and radiation era.


The amplitude of the gravitational wave with comoving wave number
$k$ remains constant when the mode lies outside the horizon.
However, once it entered the horizon, its amplitude begins to
damp. Regarding the mode which re-enters the Hubble horizon during
the matter-dominated era, we have
\cite{Boyle:2005se,Nishizawa:2017nef,Arai:2017hxj,Nunes:2018zot,Liu:2015psa,Zhao:2013bba},
\begin{equation}
    h_k^{\lambda}(\tau)=h_k^{\lambda ({\rm p})}
    \left( \frac{3j_1(k\tau)}{k\tau}\right),
\end{equation}
with $j_{\ell}$ denoting the $\ell$-th spherical Bessel function.
The Fourier-transformed tensor perturbation solution during a
cosmological era for which the Hubble rate evolves as a power-law
one, namely, $a(t)\propto t^p$, takes the form,
\begin{equation}
    h_k(\tau) \propto
    a(t)^{\frac{1-3p}{2p}}J_{\frac{3p-1}{2(1-p)}}( k\tau ),
\end{equation}
where $J_n(x)$ is the Bessel function. Another damping effect for
the $h_k(\tau)$ must be taken into account, having to do with the
fact that in the early Universe, the relativistic degrees of
freedom do not remain constant and the scale factor behaves as
$a(t) \propto T^{-1}$ \cite{Watanabe:2006qe}, so the damping
factor is,
\begin{equation}
    \left ( \frac{g_*(T_{\rm in})}{g_{*0}} \right )
    \left ( \frac{g_{*s0}}{g_{*s}(T_{\rm in})} \right )^{4/3},
\end{equation}
with $T_{\rm in}$ denoting the horizon re-entry temperature, which
is,
\begin{equation}
    T_{\rm in}\simeq 5.8\times 10^6~{\rm GeV}
    \left ( \frac{g_{*s}(T_{\rm in})}{106.75} \right )^{-1/6}
    \left ( \frac{k}{10^{14}~{\rm Mpc^{-1}}} \right ).
\end{equation}
Furthermore, another damping factor is due to the current
acceleration of the Universe $\sim (\Omega_m/\Omega_\Lambda)^2$
\cite{Boyle:2005se}. In the literature, the present day
gravitational wave energy spectrum per log frequency interval
contains the above damping effects, and also two transfer factors
that result from the numerical integration of the Fourier
transformed differential equation governing the evolution of the
tensor perturbations, so that the final expression is,
\begin{equation}
    \Omega_{\rm gw}(f)= \frac{k^2}{12H_0^2}\Delta_h^2(k),
    \label{GWspec}
\end{equation}
with $\Delta_h^2(k)$ being equal to
\cite{Boyle:2005se,Nishizawa:2017nef,Arai:2017hxj,Nunes:2018zot,Liu:2015psa,Zhao:2013bba},
\begin{equation}
\Delta_h^2(k)=\Delta_h^{({\rm p})}(k)^{2} \left (
\frac{\Omega_m}{\Omega_\Lambda} \right )^2 \left (
\frac{g_*(T_{\rm in})}{g_{*0}} \right ) \left (
\frac{g_{*s0}}{g_{*s}(T_{\rm in})} \right )^{4/3} \left
(\overline{ \frac{3j_1(k\tau_0)}{k\tau_0} } \right )^2 T_1^2\left
( x_{\rm eq} \right ) T_2^2\left ( x_R \right )\, ,
\label{mainfunctionforgravityenergyspectrum}
\end{equation}
and the ``bar'' in the Bessel function term, denoting the average
over many periods. The term $\Delta_h^{({\rm p})}(k)^{2}$ stands
for the primordial tensor power spectrum related to the
inflationary era, which is
\cite{Boyle:2005se,Nishizawa:2017nef,Arai:2017hxj,Nunes:2018zot,Liu:2015psa,Zhao:2013bba},
\begin{equation}
\Delta_h^{({\rm
p})}(k)^{2}=\mathcal{A}_T(k_{ref})\left(\frac{k}{k_{ref}}
\right)^{n_T}\, , \label{primordialtensorpowerspectrum}
\end{equation}
and it is evaluated at the CMB pivot scale
$k_{ref}=0.002$$\,$Mpc$^{-1}$, and $n_T$ is the inflationary
tensor spectral index. Here, $\mathcal{A}_T(k_{ref})$ is the
amplitude of the tensor perturbations primordially, which is
related to the amplitude of the scalar perturbations
$\mathcal{P}_{\zeta}(k_{ref})$ and the tensor-to-scalar ratio as
follows,
\begin{equation}\label{amplitudeoftensorperturbations}
\mathcal{A}_T(k_{ref})=r\mathcal{P}_{\zeta}(k_{ref})\, ,
\end{equation}
therefore the primordial tensor spectrum is finally written as
follows,
\begin{equation}\label{primordialtensorspectrum}
\Delta_h^{({\rm
p})}(k)^{2}=r\mathcal{P}_{\zeta}(k_{ref})\left(\frac{k}{k_{ref}}
\right)^{n_T}\, .
\end{equation}
Now, the transfer function $T_1(x_{\rm eq})$ in Eq.
(\ref{mainfunctionforgravityenergyspectrum}) connects basically
the gravitational wave spectrum of a mode $k$ which re-entered the
horizon near the matter-radiation equality, $t=t_{\rm eq}$ and it
is given by
\cite{Boyle:2005se,Nishizawa:2017nef,Arai:2017hxj,Nunes:2018zot,Liu:2015psa,Zhao:2013bba},
\begin{equation}
    T_1^2(x_{\rm eq})=
    \left [1+1.57x_{\rm eq} + 3.42x_{\rm eq}^2 \right ], \label{T1}
\end{equation}
with $x_{\rm eq}=k/k_{\rm eq}$ and $k_{\rm eq}\equiv a(t_{\rm
eq})H(t_{\rm eq}) = 7.1\times 10^{-2} \Omega_m h^2$ Mpc$^{-1}$.
Also the transfer function $T_2(x_R)$ in Eq.
(\ref{mainfunctionforgravityenergyspectrum}) characterizes modes
that re-entered the horizon after the reheating era commenced but
before it ended, so for $k>k_R$, and it reads,
\begin{equation}\label{transfer2}
 T_2^2\left ( x_R \right )=\left(1-0.22x^{1.5}+0.65x^2
 \right)^{-1}\, ,
\end{equation}
with $x_R=\frac{k}{k_R}$, and the $k_R$ wavenumber is,
\begin{equation}
    k_R\simeq 1.7\times 10^{13}~{\rm Mpc^{-1}}
    \left ( \frac{g_{*s}(T_R)}{106.75} \right )^{1/6}
    \left ( \frac{T_R}{10^6~{\rm GeV}} \right )\, ,  \label{k_R}
\end{equation}
where $T_R$ is the reheating temperature, and this temperature
will be a free variable in our paper, since it may affect the
present day energy spectrum of the gravitational waves. The
corresponding reheating frequency is,
\begin{equation}
    f_R\simeq 0.026~{\rm Hz}
    \left ( \frac{g_{*s}(T_R)}{106.75} \right )^{1/6}
    \left ( \frac{T_R}{10^6~{\rm GeV}} \right ),  \label{f_R}
\end{equation}
and note that the reheating frequency $f_R$ is quite close to the
most sensitive frequency band of DECIGO and BBO for $T_R \sim
10^7~{\rm GeV}$. Of course as we already mentioned, the reheating
temperature is a variable in our theory, and basically its scale
is still a mystery that refers to the hypothetical reheating era.
Finally, let us quote an expression for $g_*(T_{\mathrm{in}}(k))$
appearing in Eq. (\ref{mainfunctionforgravityenergyspectrum})
\cite{Kuroyanagi:2014nba},
\begin{align}\label{gstartin}
& g_*(T_{\mathrm{in}}(k))=g_{*0}\left(\frac{A+\tanh \left[-2.5
\log_{10}\left(\frac{k/2\pi}{2.5\times 10^{-12}\mathrm{Hz}}
\right) \right]}{A+1} \right) \left(\frac{B+\tanh \left[-2
\log_{10}\left(\frac{k/2\pi}{6\times 10^{-19}\mathrm{Hz}} \right)
\right]}{B+1} \right)\, ,
\end{align}
where $A$ and $B$ are,
\begin{equation}\label{alphacap}
A=\frac{-1-10.75/g_{*0}}{-1+10.75g_{*0}}\, ,
\end{equation}
\begin{equation}\label{betacap}
B=\frac{-1-g_{max}/10.75}{-1+g_{max}/10.75}\, ,
\end{equation}
where $g_{max}=106.75$ and $g_{*0}=3.36$. The same formulas as
above, namely Eqs. (\ref{gstartin}), (\ref{alphacap}) and
(\ref{betacap}), can be used for calculating
$g_{*0}(T_{\mathrm{in}}(k))$ by replacing $g_{*0}=3.36$ with
$g_{*s}=3.91$.

\subsection{Primordial Gravity Waves in $f(R)$ Gravity: A WKB Approach}

Let us consider the following perturbation of a flat FRW
spacetime,
\begin{equation}\label{frwmetricperturbation}
{\rm d}s^2=-a(\tau)^2{\rm d}\tau^2-2a(\tau)^2\beta_{,i}{\rm d}\tau
{\rm d}x^{i}+a(\tau)^2\left(g_{i j}^{(3)}+2\varphi g_{i j
}^{(3)}+2\gamma_{,i | j}+2h_{i j}\right ){\rm d}x^{i}{\rm d}x^{j
}\, ,
\end{equation}
with $i,j=1,2,3$, and the above perturbation includes both tensor
and scalar type perturbations, with the former being denoted as
$h_{i j}$ and it is traceless, symmetric and transverse, that is
$h_{ii}=0$, $h_{\alpha \beta}=h_{\beta \alpha}$ and
$\partial^jh_{ij}=0$ respectively. Also the scalar type
perturbations are quantified by the functions $\beta$, $\gamma$
and $\varphi$, and the spatial part of the FRW metric $g_{i
j}^{(3)}$ is,
\begin{equation}\label{spatialpartofmetric}
g_{i j}^{(3)}{\rm d}x^i{\rm d}x^j={\rm d}r^2+r^2\left({\rm
d}\theta^2+\sin^2\theta{\rm d}\phi^2 \right)\, .
\end{equation}
We shall consider two types of Lagrangian densities, and here we
shall quote two expressions for the differential equation that
governs the evolution of tensor type perturbations. Firstly,
consider a pure $f(R)$ gravity action of the form,
\begin{equation}\label{purefrlagrangian}
\mathcal{S}=\int {\rm
d}^4x\sqrt{-g}\left(\frac{f(R)}{2\kappa^2}+\mathcal{L}_m \right)\,
,
\end{equation}
where $\mathcal{L}_m$ denotes the Lagrangian density of the matter
perfect fluids that are present. The differential equation that
governs the evolution of the tensor perturbation $C_{ij}$ is the
following \cite{Hwang:2005hb},
\begin{equation}\label{evolutionequationpurefrgravity}
\frac{1}{a^3f_R}\frac{{\rm} d}{{\rm d} t}\left(a^3f_R\dot{h}_{i j}
\right)-\frac{\nabla^2}{a^2}h_{i j}=0\, ,
\end{equation}
where $\nabla^2$ is the Laplacian corresponding to the three
dimensional spacelike hypersurface spatial part of the FRW
spacetime, and $f_R=\frac{\partial f}{\partial R}$. Also we
ignored anisotropic stress effects, due to the fact that we are
interested in large frequencies, and anisotropic stress effects
are strong below $\mathcal{O}(10^{-16})$Hz, so at CMB frequencies
only. If we consider the Fourier transformation of the tensor
perturbation $h_{i j}$ given in Eq. (\ref{fourier_expand}), the
evolution equation (\ref{evolutionequationpurefrgravity}) for both
polarizations becomes,
\begin{equation}\label{fouriertransformationoftensorperturbation}
\frac{1}{a^3f_R}\frac{{\rm} d}{{\rm d} t}\left(a^3f_R\dot{h}(k)
\right)+\frac{k^2}{a^2}h(k)=0\, ,
\end{equation}
which can be rewritten as follows,
\begin{equation}\label{mainevolutiondiffeqnfrgravity}
\ddot{h}(k)+\left(3+\alpha_M
\right)H\dot{h}(k)+\frac{k^2}{a^2}h(k)=0\, ,
\end{equation}
where $\alpha_M$ for the $f(R)$ gravity case at hand is,
\begin{equation}\label{amfrgravity}
a_M=\frac{f_{RR}\dot{R}}{f_RH}\, .
\end{equation}
Note that the evolution differential equation
(\ref{mainevolutiondiffeqnfrgravity}) is valid for both the
polarization modes and in effect there is no inequivalent
propagation between these two modes. This is not however true in
the presence of a non-trivial Chern-Simons term as we show
shortly. The evolution equation
(\ref{mainevolutiondiffeqnfrgravity}) is almost identical with the
one derived in Ref.
\cite{Nishizawa:2017nef,Arai:2017hxj,Bellini:2014fua}, apart from
a typo perhaps in the second term (and of course in Ref.
\cite{Nishizawa:2017nef,Arai:2017hxj,Nunes:2018zot} the conformal
time is used), and it is identical with the one appearing in
\cite{Nunes:2018zot}. Refs.
\cite{Nishizawa:2017nef,Arai:2017hxj,Bellini:2014fua,Nunes:2018zot}
considered the more general class of Horndeski theories, but we
specified the study in $f(R)$ gravity for clarity and transparency
of the following sections. Now the study of the evolution of
primordial gravitational waves can be simplified by adopting
Nishizawa's approach \cite{Nishizawa:2017nef,Arai:2017hxj}, which
is a WKB-based method. Let us analyze it in brief, in order to
have available the formulas needed for the sections to follow.
First, let us transform the evolution equation in the conformal
time equivalent, which is,
\begin{equation}\label{mainevolutiondiffeqnfrgravityconftime}
h''(k)+\left(2+a_M \right)\mathcal{H} h'(k)+k^2h(k)=0\, ,
\end{equation}
where the prime indicates differentiation with respect to the
conformal time $\tau$  and $\mathcal{H}=\frac{a'}{a}$. Now using
Nishizawa's WKB approach, the WKB solution to the differential
equation (\ref{mainevolutiondiffeqnfrgravityconftime}) is of the
form,
\begin{equation}\label{mainsolutionwkb}
h=e^{-\mathcal{D}}h_{GR}\, ,
\end{equation}
where $h_{GR}$ is the GR waveform solution corresponding to the
differential equation
(\ref{mainevolutiondiffeqnfrgravityconftime}) with $a_M=0$ and
also $\mathcal{D}$ has the following form,
\begin{equation}\label{dform}
\mathcal{D}=\frac{1}{2}\int^{\tau}a_M\mathcal{H}{\rm
d}\tau_1=\frac{1}{2}\int_0^z\frac{a_M}{1+z'}{\rm d z'}\, ,
\end{equation}
where we expressed finally $\mathcal{D}$ as an integral over the
redshift. Note that in the case of $f(R)$ gravity, the
gravitational wave speed is equal to unity, thus an extra damping
term present in Nishizawa's paper is absent in our case. Thus in
order to find the primordial gravitational waveform in the case of
$f(R)$ gravity, one has to calculate the ``damping'' factor
$\mathcal{D}$ up to a redshift corresponding to the era suitable
for the observations of LISA, BBO, DECIGO etc. which correspond to
quite large frequencies and to modes that reentered the Hubble
horizon quite close to the reheating era. Also the ``damping''
term $e^{-\mathcal{D}}$ might not eventually be an actual damping
term since it is possible that this term might lead to an
amplification of the gravitational wave energy spectrum, due to
the specific form of the $f(R)$ gravity. This feature is more or
less model dependent, but our results indicate that most viable
$f(R)$ gravities indeed lead to damping effects, apart from the
models that we will present in this paper.

We need to note that Nishizawa's WKB solution is basically valid
only for modes that are subhorizon during present day and the
modes that reentered the Hubble horizon during the beginning of
the radiation era and during the mysterious reheating era, are
basically subhorizon modes. Let us express $a_M$ as a function of
the redshift, and we easily find that,
\begin{equation}\label{amzfr}
a_M=-\frac{f_{RR}}{f_R}(1+z)\frac{{\rm d}R(z)}{{\rm d}z}\, ,
\end{equation}
and basically in order to calculate the total ``damping'' (or
amplification factor, but we shall quote it damping factor
hereafter for convenience) factor we must integrate numerically
the Friedmann equation for $f(R)$ gravity up to a redshift that
corresponds to the era at which the modes of interest reentered
the Hubble horizon and became subhorizon modes. These modes became
subhorizon modes during the reheating era, definitely during the
radiation domination era, so the redshift is at least of the order
of $z\sim \mathcal{O}(10^6)$. In order to have a precise idea on
how large redshifts one must use in order to consistently
calculate the damping factor, recall that the redshift and the
Universe's temperature are related as follows
\cite{Garcia-Bellido:1999qrp},
\begin{equation}\label{redshiftvstemperature}
T=T_0(1+z)\, ,
\end{equation}
where $T_0$ is the present day temperature. Thus considering
temperatures corresponding to the reheating era, which are of the
GeV order, and also by taking into account that the present day
Universe's temperature is 3 Kelvin, or $T_0=2.58651\times
10^{-4}$eV, then for temperatures of the GeV order, one obtains
$z\sim 3.86621 \times 10^{12}$ or larger. It is apparent that
integrating the Friedmann equation at such large redshifts might
be a formidable task, however, things can get much easier by
recalling that when the total EoS of the Universe $\omega_{tot}$
becomes of the order of $\omega_{tot}\sim 1/3$, the scale factor
evolves as $a(t)\sim t^{1/2}$, and therefore $\dot{R}$ is
identically zero, and the same applies for the Ricci scalar $R$.
Thus, if the term $f_{RR}/f_R$ does not become singular for $R=0$
and behaves as a constant, then the parameter $a_M$ in Eq.
(\ref{amfrgravity}) becomes equal to zero. Therefore it is not
necessary to calculate the integral (\ref{dform}) for redshifts
larger than the ones for which $\omega_{tot}\sim 1/3$. This
feature greatly simplifies the calculations, since for redshifts
larger than the ones for which $\omega_{tot}\sim 1/3$, the damping
factor $e^{-\mathcal{D}}$ becomes unity, and the GR waveform
describes perfectly the gravitational wave
(\ref{mainfunctionforgravityenergyspectrum}).

\subsection{Primordial Gravity Waves in Chern-Simons $k$-essence $f(R)$ Gravity: A WKB Approach}

Apart from the pure $f(R)$ gravity models, another interesting
class of modified gravity models, that leads to particularly
interesting phenomenological predictions, is Chern-Simons
corrected $k$-essence $f(R)$ gravity. The reason we chose this
specific modified gravity to study it along with the pure $f(R)$
gravity is mainly the fact that it leads to a unique prediction
for the primordial gravitational wave energy spectrum: the
predicted signal has two components, thus two signals are
predicted and more importantly the GR signal is significantly
amplified for frequencies probed by the LISA mission, or even
higher. Consider again the perturbation of the FRW metric
appearing in Eq. (\ref{frwmetricperturbation}) and now the
underlying theory is assumed to be a Chern-Simons corrected
$k$-essence $f(R)$ gravity, in which case the general action is,
\begin{equation}\label{purefrlagrangiancs}
\mathcal{S}=\int {\rm
d}^4x\sqrt{-g}\left(\frac{f(R,X,\phi)}{2}+\mathcal{L}_m
+\frac{1}{8}\nu(\phi)\eta^{abcd}R_{ab}^{\ \ ef}R_{cdef}\right)\, ,
\end{equation}
where again $\mathcal{L}_m$ denotes the Lagrangian density of the
matter perfect fluids that are present and $\eta^{abcd}$ is a
totally antisymmetric Levi-Civita tensor density. Also $X$ in the
action above is the kinetic term
$X=\frac{1}{2}\partial^{\mu}\phi\partial_{\mu}\phi$. The action
(\ref{purefrlagrangiancs}) describes the Chern-Simons corrected
$f(R,X,\phi)$ theory with the Chern-Simons coupling being $\nu
(\phi)$. In the end, we shall specify the $f(R,X,\phi)$ theory to
be a potential-less $k$-essence $f(R)$ gravity. The Chern-Simons
term affects solely the tensor perturbations evolution, and does
not affect at all the scalar perturbations and the field equations
for a FRW spacetime. In effect, the Chern-Simons term affects the
inflationary era via the tensor-to-scalar ratio and the tensor
spectral index, and indirectly it will affect the energy spectrum
of the primordial gravitational waves via the primordial scaling
$\sim k^{n_T}$. For the action (\ref{purefrlagrangiancs}), the
evolution differential equation that governs the tensor
perturbation is \cite{Hwang:2005hb},
\begin{align}
\label{tensorpert} \frac{1}{a^3}\frac{d}{dt}\left(a^3\dot
h_{\alpha\beta}\right)-\frac{\nabla^2}{\alpha^2}h_{\alpha\beta}-\frac{2\kappa^2}{a}\epsilon_{(\alpha}^{\
\ \mu\nu}\left( (\ddot\nu-H\dot\nu)\dot h_{\beta)\mu}+\dot\nu
D_{\beta)\mu}\right)_{,\nu}=0\, ,
\end{align}
with $D_{\alpha\beta}=\ddot h_{\alpha\beta}+3H\dot
h_{\alpha\beta}-\frac{\nabla^2}{a^2}h_{\alpha\beta}$. This
evolution equation describes the tensor mode evolution, however it
is not diagonal, thus by making the following Fourier
transformation,
\begin{equation}
\centering \label{Cmodes}
h_{\alpha\beta}=\sqrt{Vol}\int{\frac{d^3k}{(2\pi)^3}\sum_{\ell}\epsilon_{\alpha\beta}^{(\ell)}(\vec{k})h_{\ell\vec{k}}e^{i\vec{k}\vec{x}}}\,
,
\end{equation}
 Eq. (\ref{tensorpert}) is diagonalized in the following way,
\begin{equation}
\centering \label{modeeq}
\frac{1}{a^3Q_t}\frac{d}{dt}\left(a^3Q_t\dot
h_{\ell\vec{k}}\right)+\frac{k^2}{a^2}h_{\ell\vec{k}}=0\, ,
\end{equation}
where,
\begin{equation}\label{qtchernsimons}
Q_t=f_R+2\lambda_{\ell}\dot\nu\frac{k}{a}
\end{equation}
and $f_R=\frac{\partial f(R,X,\phi)}{\partial R}$. In the case at
hand, $h_{\ell \vec{k}}$ describes a tensor mode perturbation with
polarization $l$, and $\epsilon_{\alpha\beta}^{(\ell)}$ denotes
the circular polarization tensor which has the property
$ik_{\gamma}\varepsilon_{\alpha}^{\gamma
\delta}\epsilon^{(\ell)}_{\beta
\delta}=k\lambda_{\ell}\epsilon^{(\ell)}_{\alpha \beta}$.
Apparently the evolution equation (\ref{modeeq}) describes the
distinct evolution of two polarization modes, the left-handed
($\lambda_L=-1$) and right handed modes ($\lambda_R=1$). The
presence of parameter $Q_t$ directly makes the propagation of
the left and right handed modes inequivalent. For the
$f(R,X,\phi)$ theory at hand, the propagation speed of the tensor
perturbations is equal to unity in natural units, thus it is equal
to the light speed, however, the wave speed of the scalar
perturbations is not equal to unity, but this will concern us when
we deal with the inflationary aspects of the current theory. For
the moment let us focus on the tensor perturbations, so the
evolution equation for a polarization $\ell$  can be written as
follows,
\begin{equation}\label{mainevolutiondiffeqnfrgravitycs}
\ddot{h}(k)+\left(3+\alpha_{M\ell}
\right)H\dot{h}(k)+\frac{k^2}{a^2}h(k)=0\, ,
\end{equation}
where $\alpha_M$ for the $f(R,X,\phi)$ gravity case at hand is
equal to,
\begin{equation}\label{amfrgravitycs}
a_{M\ell}=\frac{\dot{Q}_t}{Q_t H}\, ,
\end{equation}
and by using the definition of $Q_t$ given in
(\ref{qtchernsimons}), the parameter $a_M$ becomes,
\begin{equation}\label{amchersimonsfinal}
a_{M\ell}=\frac{f_{RR}\dot{R}+2\lambda_{\ell}\ddot{\nu}k/a-2\lambda_{\ell}\dot{\nu}kH/a}{(f_R+2\lambda_{\ell}\dot{\nu}k/a)}\,
,
\end{equation}
and note that we basically have two parameters $a_{M\ell}$, one
corresponding to the left handed and one to the right handed
polarization. The above expression can easily be expressed in
terms of the redshift, and the primordial gravitational wave form
for the theory at is again given by Eq. (\ref{dform}). Thus again
by integrating the corresponding Friedmann equation up to a
certain redshift, we may obtain the ``damping'' factor
$\mathcal{D}$ in Eq. (\ref{dform}). The new feature is that the
theory predicts basically two distinct signals of primordial
gravitational waves, corresponding to the left and right handed
polarizations which propagate in an inequivalent way. Also let us
again note that the expression for $a_{M\ell}$ is simplified for
redshifts which correspond to the radiation domination era with
$\omega_{tot}\sim 1/3$, and acquires the following form,
\begin{equation}\label{amchersimonsfinalsimpl}
a_{M\ell}=\frac{2\lambda_{\ell}\ddot{\nu}k/a-2\lambda_{\ell}\dot{\nu}kH/a}{(f_R+2\lambda_{\ell}\dot{\nu}k/a)}\,
.
\end{equation}

\subsection{Strategy for Obtaining the Precise ``Damping'' Caused by Modified Gravity Effects and Constraints}

Let us analyze in brief in this section the strategy and the aims
of the sections that follow. Our aim is to make exact predictions
for the primordial gravitational wave energy spectrum for some
$f(R)$ gravity models of particular interest. The $f(R)$ gravity
models are not randomly selected, but these satisfy quite
stringent viability criteria, with respect to their early and
late-time phenomenology. We shall consider one pure $f(R)$ gravity
model and one Chern-Simons corrected $k$-essence $f(R)$ gravity
model without scalar potential. Both these models produce a
successful inflationary era, compatible with the latest Planck
constraints on inflation \cite{Akrami:2018odb}, and we shall
extract the tensor-to-scalar ratio and the tensor spectral index
for both the models, since these two observational parameters are
essential for the calculation of the primordial gravitational wave
energy spectrum. Furthermore, the models predict a viable dark
energy era, with predictions that are compatible with the latest
Planck constraints on the cosmological parameters
\cite{Aghanim:2018eyx}. Apparently, both the models provide a
theoretical framework for which the inflationary era and the dark
energy era can be described in a unified way, thus the two models
were chosen by taking into account stringent criteria. Now with
regard to the predictions for the primordial gravitational wave
energy spectrum, our aim in both models is to calculate the
damping factor $e^{-\mathcal{D}}$, so this amounts in calculating
the integral of Eq. (\ref{dform}), namely
$\mathcal{D}=\frac{1}{2}\int_0^z\frac{a_M}{1+z'}{\rm d z'}$, from
zero redshift, which corresponds to present time, until a high
redshift corresponding to the modes that reentered the Hubble
horizon during the radiation domination era, and specifically
during the pretty much unknown reheating epoch. Most of the
proposed experiments, such as the LISA mission, BBO, DECIGO,
Einstein Telescope and so on, probe frequencies that correspond to
wavelengths of modes that reentered the Hubble horizon during the
reheating epoch, so this justifies why considering such extremely
large redshifts. Thus our central goal is to calculate the integral of
Eq. (\ref{dform}) for both models with two distinct damping effects
on the primordial gravitational waves energy spectrum, quantified by the
parameters $a_M$ appearing in Eq. (\ref{amfrgravity}) for the pure $f(R)$
gravity model, and by the parameter $a_{M\ell}$ appearing in Eq.
(\ref{amfrgravitycs}) for the Chern-Simons corrected $k$-essence
$f(R)$ gravity model. In order to calculate the integral
numerically, one must know the Hubble rate $H(z)$ for both
models, so in order to find this, we will solve numerically the
Friedmann equation for both models. To this end, we shall
introduce some statefinder variable quantities, expressed in terms
of the Hubble rate, and we shall rewrite the Friedmann equations
for both models in terms of this new statefinder quantities, using
the redshift as a dynamical variable, instead of the cosmic time.
By solving numerically the Friedmann equation, and having an
interpolating numerical solution for the Hubble rate, facilitates
very much the calculation of the damping factor for both models. A
cumbersome issue is the choice of the final redshift, since the
reheating era corresponds to significantly high redshifts. For the
pure $f(R)$ gravity case things are easy, since as we will show,
the integral beyond redshifts of the order $z\sim
\mathcal{O}(4\times 10^5)$ is essentially zero due to its highly
oscillating behavior. Apart from this model dependent feature, in
the pure $f(R)$ gravity case, when the total EoS parameter is
approximately $\omega_{tot}\sim 1/3$, the cosmological model is
described by a radiation domination era scale factor, thus $R=0$
and $\dot{R}=0$, hence for the model we shall consider, for which
$\frac{f_{RR}}{f_R}\neq 0$ for $R=0$, the parameter $a_M$ is
identically zero during the radiation domination era. This feature
can be used for future studies of pure $f(R)$ gravity models by
the readers. Also for the case of the Chern-Simons corrected
$k$-essence $f(R)$ gravity model, we shall integrate the Friedmann
equation for maximum redshifts of the order $z\sim
\mathcal{10^5}$, since beyond that the total EoS parameter becomes
approximately $\omega_{tot}\sim 1/3$ and it proves numerically
that the integral for both left and right polarization modes is
nearly zero. After calculating the exact damping for both
models, we plot the $h^2$-scaled gravitational wave energy
spectrum versus the frequency by using the $h_{GR}$ waveform with
the damping factor we calculated, and we make exact predictions
for the spectrum, comparing it with the sensitivity curves of the
most interesting future experiments.

Finally, let us bear in mind that there are upper constraints for
the gravitational wave energy spectrum, coming from various
sources, and it is worth quoting these here. Firstly, at low
multipoles of CMB and at low frequencies, of the order $f\sim
\mathcal{O}(10^{-17})\,$Hz, the gravitational wave spectrum is
constrained to be $h^2\Omega_{GW}\sim 10^{-16}$
\cite{Clarke:2020bil}, at $f\sim \mathcal{O}(3\times
10^{-16})\,$Hz, the gravitational wave spectrum is constrained to
be $h^2\Omega_{GW}< 8.4\times 10^{-7}$ \cite{Clarke:2020bil}
(without taking into account anisotropic stress effects) and for
$3\times 10^{-16}{\rm Hz}<f<10^{-15}\,$Hz, the gravitational wave
spectrum is constrained to be $h^2\Omega_{GW}< 8.6\times 10^{-7}$
(including anisotropic stress from neutrino free streaming)
\cite{Clarke:2020bil}. Also for $f\geq 10^{-15}\,$Hz, the
gravitational wave spectrum is constrained to be $h^2\Omega_{GW}<
1.7\times 10^{-6}$ for adiabatic initial conditions and
$h^2\Omega_{GW}< 2.9\times 10^{-7}$ for homogeneous initial
conditions \cite{Clarke:2020bil}. Also the Big Bang
Nucleosynthesis bound is,
\begin{equation}\label{BBNbound}
\Omega_{GW}h^2\leq \frac{n_T(2-n_T)}{2}\times 5.6\times
10^{-6}(N_{ef}-3)\, ,
\end{equation}
with $N_{ef}=4.65$ (95$\%$CL) \cite{Kuroyanagi:2014nba}.

\section{Specific Models Predictions and Quantitative Analysis}

\subsection{Unified Description of Inflation and Dark Energy I: A Pure $f(R)$ Gravity Model}

Let us first consider a pure $f(R)$ gravity model to check the
gravitational wave spectrum in this case. We have chosen an $f(R)$
gravity model which allows the unified description of the
inflationary era with the dark energy era. This model was
discussed in some previous works, see for example
\cite{Odintsov:2020nwm,Oikonomou:2020qah}, so we refer the reader
to these works for details. In this work we shall briefly discuss
the model for completeness, and in the end we shall calculate
numerically the ``damping'' factor for this model. Eventually we
shall calculate the primordial gravitational wave energy density
at present day, and we shall discuss the possibility of detection
in this case. Consider the pure $f(R)$ gravity action in the
presence of the perfect matter fluid Lagrangian $\mathcal{L}$,
which we shall assume to be cold dark matter and radiation fluids,
\begin{equation}
\label{mainaction} \mathcal{S}=\int d^4x\sqrt{-g}\left[
\frac{1}{2\kappa^2}f(R)+\mathcal{L}_m \right]\, ,
\end{equation}
where $\kappa^2=\frac{1}{8\pi G}=\frac{1}{M_p^2}$, and with $G$
being Newton's gravitational constant while $M_p$ denotes the
reduced Planck mass. The specific form of the $f(R)$ gravity that
we will use in this work is,
\begin{equation}\label{starobinsky}
f(R)=R+\frac{1}{M^2}R^2-\gamma \Lambda
\Big{(}\frac{R}{3m_s^2}\Big{)}^{\delta}\, ,
\end{equation}
with $m_s$ in Eq. (\ref{starobinsky}) being
$m_s^2=\frac{\kappa^2\rho_m^{(0)}}{3}$, and $\rho_m^{(0)}$ being
the energy density of cold dark matter today. Parameter
$\delta $ is deliberately chosen to take values in the interval
$0<\delta <1$, $\gamma$ is a freely chosen dimensionless
parameter, and parameter $\Lambda $ is a free parameter with
mass dimensions $[m]^2$. Furthermore, for phenomenological reasons
\cite{Appleby:2009uf}, parameter $M$ is chosen to be $M=
1.5\times 10^{-5}\left(\frac{N}{50}\right)^{-1}M_p$, where $N$ is
number of  $e$-foldings during inflation. For a flat FRW
spacetime, upon varying the action (\ref{mainaction}) with respect
to the metric, we get the following equations of motion,
\begin{align}\label{eqnsofmkotion}
& 3 H^2f_R=\frac{Rf_R-f}{2}-3H\dot{f}_R+\kappa^2\left(\rho_m+
\rho_r\right)\, ,\\ \notag & -2\dot{H}f_R=\ddot{f}_R-H\dot{f}_R
+\frac{4\kappa^2}{3}\rho_r\, ,
\end{align}
where $f_R=\frac{\partial f}{\partial R}$ and the ``dot'' as usual
denotes differentiation with respect to the cosmic time. Also
$\rho_r$ and $\rho_m$ denote the radiation and cold dark matter
energy densities respectively. The model (\ref{starobinsky})
produces an inflationary era which is basically controlled by the
$R^2$ model, as we now evince. Apparently the $\sim R^{\delta}$
term is subleading at early times, as we explicitly show now, and
controls the late-time dynamics. For the model
(\ref{starobinsky}), the equations of motion (\ref{eqnsofmkotion})
become,
\begin{equation}\label{friedmanequationinflation}
3H^2\left(1+\frac{2}{M^2}R-
\frac{\gamma\delta\Lambda}{(3m_s^2)^\delta}R^{\delta-1}\right)=\frac{R^2}{2M}+\frac{\gamma(1-
\delta)\Lambda}{2}
\Big{(}\frac{R}{3m_s^2}\Big{)}^{\delta}-3H\dot{R}\Big{(}\frac{2}{M^2}-\frac{\gamma
\delta (\delta-1)\Lambda}{(3m_s^2)^\delta}R^{\delta-2}\Big{)}+
\kappa^2\Big{(}\rho_r+\rho_m \Big{)}\, .
\end{equation}
We shall assign the following values for the free dimensionless
parameters $\gamma $ and $\delta $,
\begin{equation}\label{gammaanddelta}
\gamma=\frac{1}{0.5},\,\,\, \delta=\frac{1}{100}\, ,
\end{equation}
and moreover let $\Lambda\simeq 11.895\times 10^{-67}$eV$^2$ which
is chosen to take values almost identical with the present day
cosmological constant. Also recall that $m_s^2\simeq 1.87101\times
10^{-67}$eV$^2$ and  $M= 1.5\times
10^{-5}\left(\frac{N}{50}\right)^{-1}M_p$ \cite{Appleby:2009uf},
so for $N\sim 60$, $M$ is $M\simeq 3.04375\times 10^{22}$eV.
Assuming a slow-roll evolution during inflation, we have $R\simeq
12 H^2$, so for a low-scale inflationary era, with $H= H_I\sim
10^{13}$GeV, the curvature scalar becomes approximately $R\sim
1.2\times 10^{45}$eV$^2$. Having these data at hand, we can
proceed and directly compare the terms of Eq.
(\ref{friedmanequationinflation}),

For a low-scale inflationary era, with $H= H_I\sim 10^{13}$GeV, we
have $R\sim 1.2\times \mathcal{O}(10^{45})$eV$^2$, $R^2/M^2\sim
\mathcal{O}(1.55\times
 10^{45})$eV$^2$ and also $\sim \Big{(}\frac{R}{3m_s^2}\Big{)}^{\delta}\sim
 \mathcal{O}(10)$. Moreover we have $\sim \Big{(}\frac{R}{3m_s^2}\Big{)}^{\delta-1}\sim
 \mathcal{O}(10^{-111})$ and lastly $\sim \Big{(}\frac{R}{3m_s^2}\Big{)}^{\delta-2}\sim
 \mathcal{O}(10^{-223})$. Therefore only the terms with positive powers of
 the curvature dominate the Friedmann equation
 (\ref{friedmanequationinflation}), which becomes at leading
 order,
\begin{equation}\label{friedmanequationinflationaux}
3H^2\left(1+\frac{2}{M^2}R\right)=\frac{R^2}{2M}-\frac{6H\dot{R}}{M^2}\,
,
\end{equation}
which is basically identical with the $R^2$ model case. Rewriting
it we get,
\begin{equation}\label{patsunappendixinflation}
3\ddot{H}-3\frac{\dot{H}^2}{H}+\frac{2M^2H}{6}=-9H\dot{H}\, ,
\end{equation}
and by solving it, assuming a slow-roll evolution we have,
\begin{equation}\label{quasidesitter}
H(t)=H_0-\frac{M^2}{36} t\, ,
\end{equation}
which is a quasi-de Sitter evolution that leads directly to the
$R^2$ model spectral index and tensor-to-scalar ratio, which are,
\begin{equation}\label{r2modelspectralindices}
n_s\sim 1-\frac{2}{N},\,\,\,r\sim \frac{12}{N^2}\, .
\end{equation}
Also let us calculate in detail the tensor-spectral index for the
$R^2$ model, which is essential for the calculation of the
primordial gravitational wave energy spectrum. Recall the
slow-roll indices for $f(R)$ gravity are
\cite{Hwang:2005hb,reviews1,Odintsov:2020thl},
\begin{equation}
\label{restofparametersfr}\epsilon_1=-\frac{\dot{H}}{H^2}, \quad
\epsilon_2=0\, ,\quad \epsilon_3= \frac{\dot{f}_R}{2Hf_R}\, ,\quad
\epsilon_4=\frac{\ddot{f}_R}{H\dot{f}_R}\,
 ,
\end{equation}
and from the Raychaudhuri equation for $f(R)$ gravity we have,
\begin{equation}\label{approx1}
\epsilon_1=-\epsilon_3(1-\epsilon_4)\, .
\end{equation}
The tensor spectral index for $f(R)$ gravity is equal to
\cite{Hwang:2005hb,reviews1,Odintsov:2020thl},
\begin{equation}\label{tensorspectralindexr2gravity}
n_T\simeq -2 (\epsilon_1+\epsilon_3)\, ,
\end{equation}
so in order to compute it for the $R^2$ gravity, we need an
approximate expression for $\epsilon_4$. Let us focus on the
slow-roll index $\epsilon_4$, and we easily find that,
\begin{equation}\label{epsilon41}
\epsilon_4=\frac{\ddot{f}_R}{H\dot{f}_R}=\frac{\frac{d}{d
t}\left(f_{RR}\dot{R}\right)}{Hf_{RR}\dot{R}}=\frac{f_{RRR}\dot{R}^2+f_{RR}\frac{d
(\dot{R})}{d t}}{Hf_{RR}\dot{R}}\, .
\end{equation}
However, $\dot{R}$ is equal to,
\begin{equation}\label{rdot}
\dot{R}=24\dot{H}H+6\ddot{H}\simeq 24H\dot{H}=-24H^3\epsilon_1\, ,
\end{equation}
where  the slow-roll approximation $\ddot{H}\ll H \dot{H}$ is used
above and also we assumed in Eq. (\ref{rdot}) that $\dot{R}\simeq
24H\dot{H}\simeq -24H^3\epsilon_1$, hence we avoided the
appearance of the term $\sim \frac{d^3H}{dt^2}$ in the final
expression. Combining Eqs. (\ref{rdot}) and (\ref{epsilon41})
after some algebra we get,
\begin{equation}\label{epsilon4final}
\epsilon_4\simeq -\frac{24
f_{RRR}H^2}{f_{RR}}\epsilon_1-3\epsilon_1+\frac{\dot{\epsilon}_1}{H\epsilon_1}\,
,
\end{equation}
but $\dot{\epsilon}_1$ is explicitly equal to,
\begin{equation}\label{epsilon1newfiles}
\dot{\epsilon}_1=-\frac{\ddot{H}H^2-2\dot{H}^2H}{H^4}=-\frac{\ddot{H}}{H^2}+\frac{2\dot{H}^2}{H^3}\simeq
2H \epsilon_1^2\, ,
\end{equation}
hence the final approximate expression for $\epsilon_4$ is,
\begin{equation}\label{finalapproxepsilon4}
\epsilon_4\simeq -\frac{24
f_{RRR}H^2}{F_{RR}}\epsilon_1-\epsilon_1\, .
\end{equation}
For the case of $R^2$ gravity obviously $\epsilon_4\simeq
-\epsilon_1$, hence the tensor spectral index for the $R^2$ model
reads,
\begin{equation}\label{tensorspectralindexr2ini}
n_T\simeq -2 \frac{\epsilon_1^2}{1+\epsilon_1}\simeq
-2\epsilon_1^2\, ,
\end{equation}
and due to the fact that $\epsilon_1\simeq \frac{1}{2N^2}$ for the
$R^2$ model, the resulting expression for the tensor spectral
index corresponding to the $R^2$ model is,
\begin{equation}\label{r2modeltensorspectralindexfinal}
n_T\simeq -\frac{1}{2N^2}\, ,
\end{equation}
so for $N=60$ it is approximately $n_T=-0.000138889$ and it is
always red-tilted of course. In the same vain, for $N=60$, the
tensor-to-scalar ratio $r=12/N^2$ is equal to $r=0.00333333$ and
we shall use both these values for the calculation of the
gravitational wave energy spectrum. Having dealt with the
early-time era essentials, let us proceed in finding a way to
express the field equations in terms of appropriate statefinder
variables and in terms of the redshift. This will make the
integration needed for the calculation of the damping factor
easier, since we shall basically numerically integrate the
statefinder transformed equations of motion. First let us note
that we can form the $f(R)$ gravity equations of motion to have
the most familiar Einstein-Hilbert form as follows,
\begin{align}\label{flat}
& 3H^2=\kappa^2\rho_{tot}\, ,\\ \notag &
-2\dot{H}=\kappa^2(\rho_{tot}+P_{tot})\, ,
\end{align}
where $\rho_{tot}=\rho_{m}+\rho_{DE}+\rho_r$ denotes the total
energy density of the cosmological fluid and $P_{tot}=P_r+P_{DE}$
denotes the total pressure. The total fluid has three components,
the cold dark matter, the pressure and the geometric fluid, which
at early times drives inflation, and at late times drives the dark
energy era. The energy density of the geometric fluid reads,
\begin{equation}\label{degeometricfluid}
\rho_{DE}=\frac{f_R R-f}{2}+3H^2(1-f_R)-3H\dot{f}_R\, ,
\end{equation}
and its pressure reads,
\begin{equation}\label{pressuregeometry}
P_{DE}=\ddot{f}_R-H\dot{f}_R+2\dot{H}(f_R-1)-\rho_G\, .
\end{equation}
All the fluids considered are perfect fluids, and satisfy the
following continuity equations,
\begin{align}\label{fluidcontinuityequations}
& \dot{\rho}_m+3H(\rho_m+P_m)=0\, , \\ \notag &
\dot{\rho}_r+3H(\rho_r+P_r)=0\, , \\ \notag &
\dot{\rho}_{DE}+3H(\rho_{DE}+P_{DE})=0\, .
\end{align}
Let us use the redshift $z$ as a dynamical variable, defined as
follows,
\begin{equation}\label{redshift}
1+z=\frac{1}{a}\, ,
\end{equation}
where the present day scale factor $a_0$, thus at  $z=0$, is
assumed to be unity for the simple reason that the physical and
comoving wavenumbers at present time, $k$ and $a_0k$ coincide. We
introduce the following statefinder quantity
\cite{Bamba:2012qi,Odintsov:2020nwm,Oikonomou:2020qah}, follows,
\begin{equation}\label{yHdefinition}
y_H(z)=\frac{\rho_{G}}{\rho^{(0)}_m}\, ,
\end{equation}
where $\rho^{(0)}_m$ is the energy density of cold dark matter at
present time. We can express the equations of motion in terms of
the statefinder $y_H(z)$, by using the Friedmann equation, and we
have,
\begin{equation}\label{yhfunctionanalyticzero}
y_H(z)=\frac{3H^2}{\kappa^2\rho^{(0)}_m}-\frac{\rho_m}{\rho^{(0)}_m}-\frac{\rho_r}{\rho^{(0)}_m}\,
.
\end{equation}
But $\rho_r$ redshifts as $\rho_r=\rho_r^{(0)}a^{-4}$, with
$\rho_r^{(0)}$ being the value of the radiation energy density
today, so $\frac{\rho_r}{\rho^{(0)}_m}=\chi (1+z)^4$, where
$\chi=\frac{\rho^{(0)}_r}{\rho^{(0)}_m}\simeq 3.1\times 10^{-4}$.
Also $\rho_m$ redshifts as $\rho_m=\rho_m^{(0)}a^{-3}$ thus, the
function $y_H(z)$ of Eq. (\ref{yhfunctionanalyticzero}) can be
written as follows,
\begin{equation}\label{finalexpressionyHz}
y_H(z)=\frac{H^2}{m_s^2}-(1+z)^{3}-\chi (1+z)^4\, .
\end{equation}
with
$m_s^2=\frac{\kappa^2\rho^{(0)}_m}{3}=H_0\Omega_c=1.37201\times
10^{-67}$eV$^2$. By using the function $y_H(z)$ we can rewrite the
cosmological equations of motion as one equation
\cite{Bamba:2012qi,Odintsov:2020nwm,Oikonomou:2020qah},
\begin{equation}\label{differentialequationmain}
\frac{d^2y_H(z)}{d z^2}+J_1\frac{d y_H(z)}{d z}+J_2y_H(z)+J_3=0\,
,
\end{equation}
with the functions $J_1$, $J_2$ and $J_3$ being defined in the
following way,
\begin{align}\label{diffequation}
& J_1=\frac{1}{z+1}\left(
-3-\frac{1-f_R}{\left(y_H(z)+(z+1)^3+\chi (1+z)^4\right) 6
m_s^2f_{RR}} \right)\, , \\ \notag & J_2=\frac{1}{(z+1)^2}\left(
\frac{2-f_R}{\left(y_H(z)+(z+1)^3+\chi (1+z)^4\right) 3
m_s^2f_{RR}} \right)\, ,\\ \notag & J_3=-3(z+1)-\frac{\left(1-f_R
\right)\Big{(}(z+1)^3+2\chi (1+z)^4
\Big{)}+\frac{R-f}{3m_s^2}}{(1+z)^2\Big{(}y_H(z)+(1+z)^3+\chi(1+z)^4\Big{)}6m_s^2f_{RR}}\,
,
\end{align}
with $f_{RR}=\frac{\partial^2 f}{\partial R^2}$. The central point
of our analysis is to solve numerically the above equation
(\ref{differentialequationmain}) by choosing suitable initial
conditions, from redshift zero, up to a suitably high redshift
that will be determined by the phenomenology of the model and the
behavior of the parameter $a_M$ in Eq. (\ref{amzfr}). Then by
finding the numerical solution $y_H(z)$, we can easily express all
the necessary quantities needed for  the calculation of $a_M$, as
functions of $y_H(z)$. Particularly, the Ricci scalar is written
in terms of $y_H(z)$ as follows,
\begin{equation}\label{ricciscalarasfunctionofz}
R(z)=3m_s^2\left( 4y_H(z)-(z+1)\frac{d y_H(z)}{d
z}+(z+1)^3\right)\, .
\end{equation}
Furthermore, the Hubble rate in terms of $y_H(z)$ is written as,
\begin{equation}\label{lambdacdmhubblerate}
H_{\Lambda}(z)=H_0\sqrt{\Omega_{\Lambda}+\Omega_M(z+1)^3+\Omega_r(1+z)^4}\,
,
\end{equation}
where $H_0$ is the value of the Hubble rate today which is
$H_0\simeq 1.37187\times 10^{-33}$eV based on the latest Planck
data \cite{Aghanim:2018eyx}, also $\Omega_{\Lambda}\simeq
0.681369$ and $\Omega_M\sim 0.3153$ \cite{Aghanim:2018eyx}, while
$\Omega_r/\Omega_M\simeq \chi$, and $\chi$ was defined earlier in
this section. Moreover, the dark energy density parameter
$\Omega_{DE}$ reads,
\begin{equation}\label{omegaglarge}
\Omega_{DE}(z)=\frac{y_H(z)}{y_H(z)+(z+1)^3+\chi (z+1)^4}\, ,
\end{equation}
and the dark energy EoS parameter reads,
\begin{equation}\label{omegade}
\omega_{DE}(z)=-1+\frac{1}{3}(z+1)\frac{1}{y_H(z)}\frac{d
y_H(z)}{d z}\, .
\end{equation}
Finally, the deceleration parameter and the total EoS parameter as
functions of the redshift and $y_{H}(z)$ read,
\begin{align}\label{statefinders}
& q=-1-\frac{\dot{H}}{H^2}=\frac{(z+1) H'(z)}{H(z)}-1\, ,
\end{align}
\begin{equation}\label{totaleosparameter}
\omega_{tot}=-1-\frac{2\dot{H}}{3H^2}=-1+\frac{2(1+z)H'(z)}{3H(z)}\, .
\end{equation}
The initial conditions we shall use were initially ``engineered''
to fit in integrations for which the  redshifts belonged deeply in
the matter domination era, however as we realized by analyzing the
results, these do not significantly affect the results, at least
in the context of pure $f(R)$ gravity. In the next section however
we shall use another set of initial conditions, different from
what we will use for the pure $f(R)$ gravity, but there the
problem and theory are different from the pure $f(R)$ gravity
case. A nice working project is to employ cosmographic arguments
\cite{Benetti:2019gmo} in order to determine the most appropriate
set of initial conditions, and this issue should be appropriately
addressed in a future work. The set of initial conditions we shall
use in this section is the following,
\begin{equation}\label{generalinitialconditions}
y_H(z_f)=\frac{\Lambda}{3m_s^2}\left(
1+\frac{(1+z_f)}{1000}\right)\, , \,\,\,\frac{d y_H(z)}{d
z}\Big{|}_{z=z_f}=\frac{1}{1000}\frac{\Lambda}{3m_s^2}\, ,
\end{equation}
where $z_f$ is the final redshift value and we shall discuss its
value now. We examined several values for the final redshift, and
as our analysis indicated, and we shall discuss shortly
quantitatively by using numerical analysis arguments, the final
redshift can be chosen to be $z_f=1000$. The reason for this is
that beyond that redshift, the integral of $a_M$ which is
basically the damping factor (\ref{dform}), does not receive any
contribution beyond $z_f=1000$ since the integral is basically
zero due to the oscillating behavior. We will support this
graphically shortly. Before going into that, let us demonstrate
the viability of the dark energy era for the model at hand by
solving numerically the differential equation
(\ref{differentialequationmain}) by using the initial conditions
(\ref{generalinitialconditions}) for $z_f=1000$, and also we shall
compare the present $f(R)$ gravity model with the $\Lambda$CDM
model. Let us now present the behavior of the cosmological
parameters for the $f(R)$ gravity model, and after this brief
analysis we proceed to the calculation of the parameter $a_M$ and
the calculation of the damping factor integral. We start off with
the behavior of the statefinder $y_H$ as a function of the
redshift with the result of our numerical analysis being presented
in Fig. \ref{plot1}. As it can be seen, near the present epoch the
statefinder approaches a constant value, but it is mentionable to
state that the dark energy era for the $f(R)$ gravity model is
described by a dynamical dark energy era, and it is not similar to
a cosmological constant, in which case the statefinder $y_H$ would
be a constant and it would not have any variation as the redshift
changes.
\begin{figure}[h!]
\centering
\includegraphics[width=25pc]{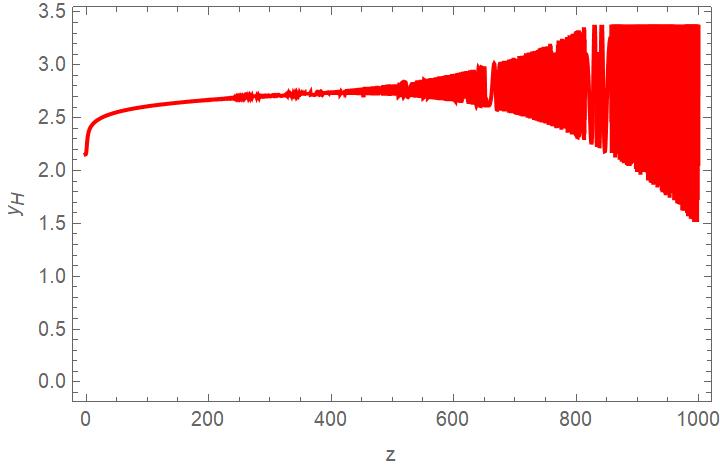}
\caption{The statefinder function $y_H$ versus the redshift.}
\label{plot1}
\end{figure}
From Fig. \ref{plot1} it is apparent that the statefinder $y_H$
strongly oscillates, a features that it seems to be
model-dependent though. Let us also present the behavior of the
dark energy EoS parameter $\omega_{DE}$ as a function of the
redshift and it is presented in Fig. \ref{plot2}.
\begin{figure}[h!]
\centering
\includegraphics[width=20pc]{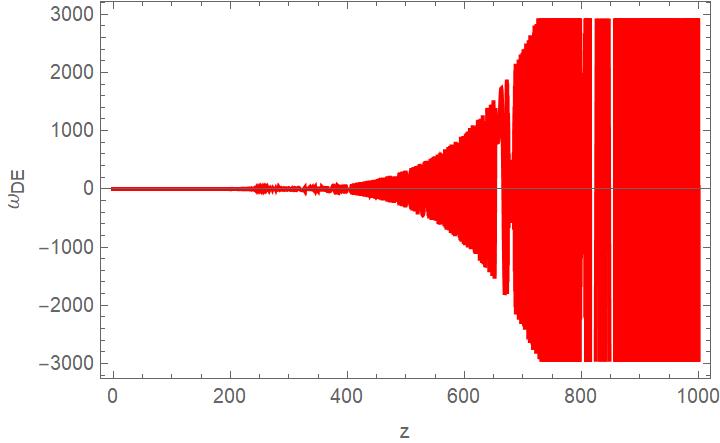}
\includegraphics[width=20pc]{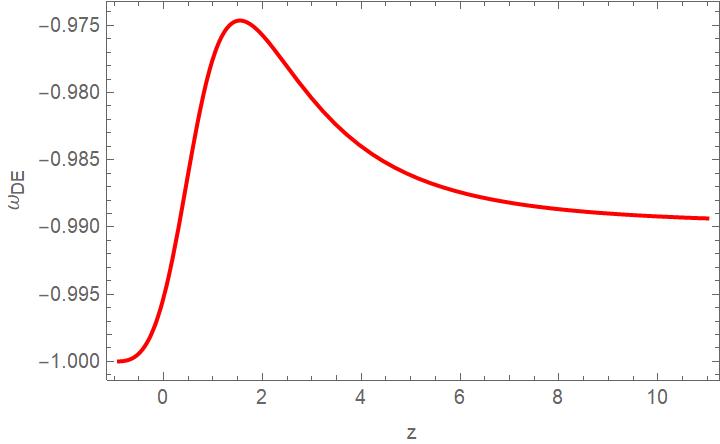}
\caption{The dark energy EoS parameter $\omega_{DE}(z)$ for
$z=[0,1000]$(left plot) and $z=[0,11]$ (right plot).}
\label{plot2}
\end{figure}
From Fig. \ref{plot2} it is also apparent that the dark energy era
for this $f(R)$ gravity model is totally different from a simple
cosmological constant since the dark energy EoS parameter is
dynamically evolving. It is also very important to see how the
total EoS parameter evolves for the $f(R)$ gravity model. The
total EoS parameter will indicate how the model evolves during the
dark energy era, the accordingly matter dominated era and how it
approaches the radiation domination era. In Fig. \ref{plot3} we
present the plots of the total EoS parameter for the $f(R)$
gravity model (red curve) compared with the $\Lambda$CDM model
(black dashed curve). As it can be seen, the model evolves from an
accelerating era to the matter domination era, with the total EoS
parameter value increasing gradually. At $z_{fin}=1000$ the value
of the total EoS parameter for the $f(R)$ gravity model is
$\omega_{tot}(1000)=0.0789406$, hence the pure matter domination
era has passed and the model starts to approach the radiation
domination era.
\begin{figure}[h!]
\centering
\includegraphics[width=25pc]{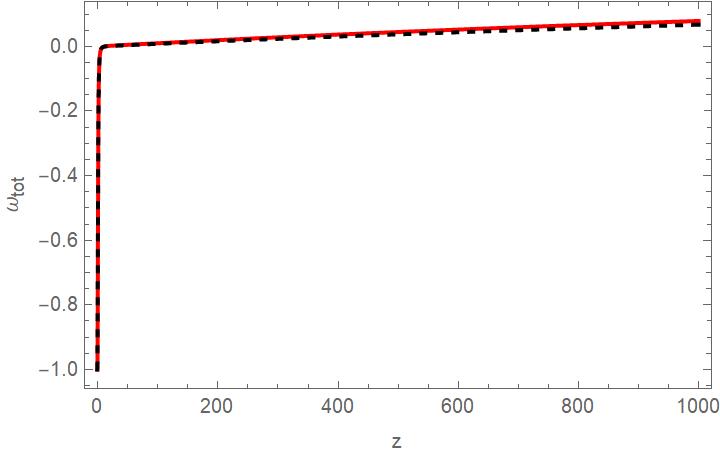}
\caption{The total EoS parameter $\omega_{tot}(z)$ for the $f(R)$
gravity model (red curve) and the $\Lambda$CDM model (black dashed
curve).} \label{plot3}
\end{figure}
Finally, in Fig. \ref{plot4} we plot the behavior of the
deceleration parameter $q$ as a function of the redshift for the
$f(R)$ gravity model (red curve) and the $\Lambda$CDM model (black
dashed curve). As it can be seen the $f(R)$ gravity model and the
$\Lambda$CDM model are indistinguishable even for such high
redshifts.
\begin{figure}[h!]
\centering
\includegraphics[width=25pc]{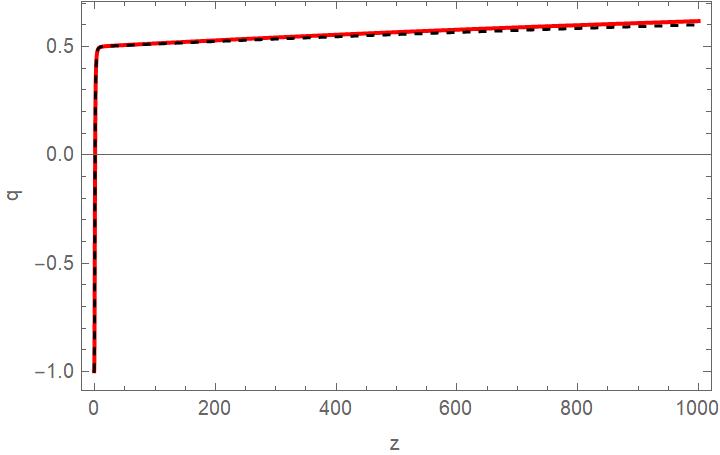}
\caption{The deceleration parameter $q$ for the $f(R)$ gravity
model (red curve) and the $\Lambda$CDM model (black dashed
curve).} \label{plot4}
\end{figure}
Also in Table \ref{table1} we present the values of the dark
energy EoS parameter and of the dark energy density parameter at
present day and we compare these with the latest Planck
constraints \cite{Aghanim:2018eyx}. Also we quote the value of the
deceleration parameter at present day for the $f(R)$ gravity model
and the $\Lambda$CDM model.
\begin{table}[h!]
  \begin{center}
    \caption{\emph{\textbf{Values of Cosmological Parameters for $f(R)$ Gravity Model and $\Lambda$CDM Model.}}}
    \label{table1}
    \begin{tabular}{r|r|r}
     \hline
      \textbf{Cosmological Parameter} & \textbf{$f(R)$ Gravity Value} & \textbf{Base $\Lambda$CDM or Planck 2018 Value} \\
           \hline
      $\Omega_{DE}(0)$ & 0.683951 & $0.6847\pm 0.0073$ \\ \hline
      $\omega_{DE}(0)$ & -0.995258 & $-1.018\pm 0.031$\\ \hline
      $q(0)$ & -0.521012 & -0.535\\
      \hline
    \end{tabular}
  \end{center}
\end{table}
The whole analysis indicates that the dark energy era and the
epochs before it for the $f(R)$ gravity model are quite successful
and mimic to a great extend the $\Lambda$CDM model, with the
difference being that the statefinder $y_H$ is dynamical for the
$f(R)$ gravity model and the dark energy EoS parameter is not
constant, but dynamically evolving.

\begin{figure}[h!]
\centering
\includegraphics[width=25pc]{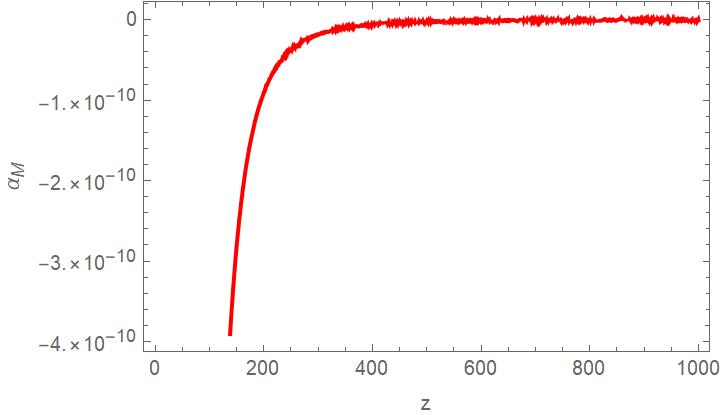}
\caption{The parameter $a_M$ (\ref{amzfr}) for the $f(R)$ gravity
model for $z=[0,1000]$.} \label{plot5}
\end{figure}

\begin{figure}[h!]
\centering
\includegraphics[width=40pc]{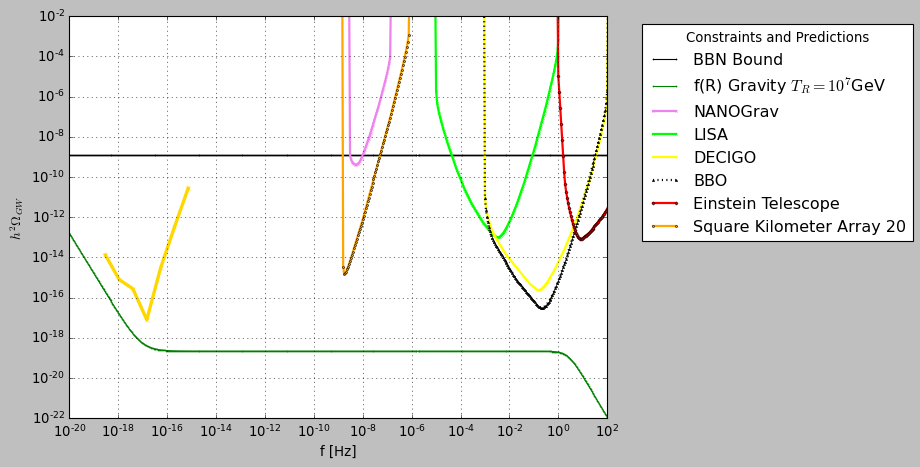}
\caption{The $h^2$-scaled gravitational wave energy spectrum for
pure $f(R)$ gravity.} \label{plotfinalfrpure}
\end{figure}

Let us now proceed to the core of our analysis for the $f(R)$
gravity model at hand, and the calculation of the parameter $a_M$
(\ref{amzfr}), the ``damping'' factor $\mathcal{D}$ (\ref{dform})
and subsequently the gravitational wave energy spectrum
(\ref{GWspec}). In Fig. \ref{plot5} we plot the parameter $a_M$
(\ref{amzfr}) as a function of the redshift. It is apparent that
beyond $z=1000$ the parameter $a_M$ oscillates and thus the
integral (\ref{dform}) receives insignificant contribution beyond
$z=1000$, a fact that we verified numerically by solving the
equations numerically for up to $z=10000$.
Thus performing the integral (\ref{dform}) numerically, we obtain
$\mathcal{D}=-0.00451304/2$ therefore the total ``damping'' factor
is for the model at hand
$e^{-2\mathcal{D}}=e^{0.00451304}=1.00452$ for redshifts up to
$z_{fin}=1000$. Also for redshifts corresponding to the radiation
domination era, the parameter $a_M$ is equal to zero since
$\dot{R}=0$ for $\omega_{tot}=1/3$ and also the term $f_{RR}/f_R$
for $R\to 0$ and for the model at hand is zero. Thus the only
contribution to the integral (\ref{dform}) is from the interval
$z=[0,1000]$, hence for this particular model, the GR waveform is
almost identical with the $f(R)$ gravity waveform. Therefore, the
gravitational wave spectrum for the $f(R)$ gravity model at hand
is,
\begin{align}
\label{GWspecfR}
    &\Omega_{\rm gw}(f)=1.00452\times \frac{k^2}{12H_0^2}r\mathcal{P}_{\zeta}(k_{ref})\left(\frac{k}{k_{ref}}
\right)^{n_T}\times \\ \notag & \left (
\frac{\Omega_m}{\Omega_\Lambda} \right )^2
    \left ( \frac{g_*(T_{\rm in})}{g_{*0}} \right )
    \left ( \frac{g_{*s0}}{g_{*s}(T_{\rm in})} \right )^{4/3} \nonumber  \left (\overline{ \frac{3j_1(k\tau_0)}{k\tau_0} } \right )^2
    T_1^2\left ( x_{\rm eq} \right )
    T_2^2\left ( x_R \right )
\end{align}
where the parameters appearing in Eq. (\ref{GWspecfR}) are defined
below Eq. (\ref{GWspec}). Now since $\tau_0=14379.2$$\,$Mpc, and
by taking a reheating temperature of the order $T_R\sim 10^7$GeV,
in Fig. \ref{plotfinalfrpure} we plot the $h^2$-scaled
gravitational wave spectrum taking also into account that
$\mathcal{P}_{\zeta}(k_{ref})\simeq 20.0855 \times 10^{-10}$ for
the latest Planck constraints on inflation \cite{Akrami:2018odb}.
Apparently, the results obtained from Fig. \ref{plotfinalfrpure}
are strict and rigid. The signal of the gravitational wave
spectrum for the model of $f(R)$ gravity we discussed in this
section, cannot be detected by any of the future experiments and
missions. Thus this offers a perspective for future studies: If a
signal is not detected by LISA or Litebird or BBO and DECIGO, this
would not necessarily mean that inflation never took place, but
possibly that the sensitivities of the experiments are not
adequate to capture this signal. We tried a large number of pure
$f(R)$ gravity models which we chose not to present and the result
seems to be robust for all the models, the signal is lower
compared to the sensitivities of the experiments. The reheating
temperature seems not to affect the result and the same applies
for the initial conditions we chosen to numerically integrate the
Friedmann equation for large redshifts. The only thing that seems
to crucially affect the resulting signal is the tensor spectral
index, and specifically if it is positive, then the signal of the
primordial gravitational waves spectrum might be detectable by the
future experiments. However a positive spectral index cannot be
obtained for $f(R)$ gravity as it is apparent from Eq.
(\ref{tensorspectralindexr2ini}). Hence in order to obtain a
positive spectral index for $f(R)$ gravity, one needs to include
other modified gravity terms in the inflationary Lagrangian. A
model of this sort is presented in the next section.

\subsection{Unified Description of Inflation and Dark Energy II: A Chern-Simons Corrected $k$-essence $f(R)$ Gravity Model}

In this section we shall consider another interesting toy model
which has two unique characteristics, firstly it describes in a
unified way and successfully both the inflationary era and the
late-time era, and secondly and more importantly it generates
detectable gravitational wave energy spectrum. Also the generated
spectrum has two distinct signals, which is due to the presence of
a Chern-Simons term. The toy model is a Chern-Simons $k$-essence
$f(R)$ gravity theory, and the motivation for using such an
extension was mainly the production of a blue tilted tensor
spectral index. As we saw in the previous section $f(R)$ gravity
produces a red-tilted tensor spectral index, thus an extension of
pure $f(R)$ gravity is needed in order to produce a blue-tilted
tensor spectral index. Apart from the possibility of producing a
blue-tilted tensor spectral index, the model eventually produces
two distinct signals in the energy spectrum of the primordial
gravitational waves, which is a quite intriguing feature and a
prediction for future predictions.

Let us commence with the specification of the gravitational
action. As we already mentioned, we shall examine the
phenomenological implications of a potential-less $k$-essence
$f(R)$ gravity in the presence of a Chern-Simons term
corresponding to the following action,
\begin{equation}
\centering
\label{action}
S=\int{d^4x\sqrt{-g}\left(\frac{f(R)}{2\kappa^2}+G(X)+\frac{1}{8}\nu(\phi)\eta^{abcd}R_{ab}^{\ \ ef}R_{cdef}\right)}\, ,
\end{equation}
with $f(R)$ being an arbitrary for the time being function
depending solely on the Ricci scalar, $\kappa=\frac{1}{M_P}$
stands for the gravitational constant with $M_P$ being the reduced
Planck mass, $G(X)$ is also an arbitrary function that depends on
the kinetic term of the scalar field
$X=\frac{1}{2}\nabla_\mu\phi\nabla^\mu\phi$, $\nu(\phi)$ signifies
the Chern-Simons scalar coupling function coupled to the
antisymmetric term. For the purpose of this paper we shall limit
our work to the case of a homogeneous and isotropic expansion
corresponding to a FRW metric. As a consequence, it is reasonable
to assume that the scalar field itself is also homogeneous,
therefore the kinetic term is simplified significantly. For the
sake of consistency we mention that hereafter the previously
introduced functions and auxiliary parameters have the following
form,
\begin{equation}
\centering \label{latetimef2initial} f(R)=R+\alpha
R^2-(R+R_0)P_3\left(\frac{M_P^2}{R+R_0}\right)\, ,
\end{equation}
with $\alpha=\frac{1}{36H_i}$ and,
\begin{align}
\centering \label{auxiliary1}G(X)=-X-f_1\kappa^4
X^2,\,\,\,X=-\frac{1}{2}\dot\phi^2,\,\,\,\phi(t)=\frac{t}{\kappa^2f_1^{\frac{1}{3}}}\,
,
\end{align}
where the parameter $H_i$ has mass dimensions $[H_i]=$eV$^2$, and
$P_3(x)$ in Eq. (\ref{latetimef2initial}) is the Legendre
polynomial of degree three. Note that in Eq. (\ref{auxiliary1})
the expression of the scalar field is taken from Ref.
\cite{Nojiri:2019dqc} for the case of $k$-Essence models during
the inflationary era. The corresponding continuity equation for
the scalar field reads,
\begin{equation}
\centering
\label{conteq1}
3f_12^{1-m}mH\dot\phi^{2m-1}-3H\dot\phi-\ddot\phi f_12^{2-m}m^2\dot\phi^2-\ddot\phi-\dot\phi^{2m-4}-f_1^22^{-m}m\dot\phi^{2m-2}\ddot\phi+2^{-m}m\ddot\phi\dot\phi^{2m-2}=0\, ,
\end{equation}
therefore by keeping the leading order terms in the aforementioned
equation, one finds,
\begin{equation}
\centering
\label{conteq2}
3f_12^{1-m}mH\dot\phi^{2m}-3H\dot\phi=0\, ,
\end{equation}
from which the solution for $\phi(t)$ appearing in Eq.
(\ref{auxiliary1}) emerges for the case of $m=2$.

As it can easily be seen, the Legendre polynomial related term of
the $f(R)$ gravity in the large curvature limit, which basically
describes inflation, behaves as follows,
\begin{equation}\label{legendrepolynomiallargecurvlimit}
\lim_{R\to
\infty}(R+R_0)P_3\left(\frac{M_P^2}{R+R_0}\right)\simeq\frac{15
M_P^6 R_0^2}{2 R^4}-\frac{5 \left(M_P^6 R_0\right)}{R^3}+\frac{5
M_P^6}{2 R^2}-\frac{3 M_P^2}{2}+\mathcal{O}(1/R^5)\, ,
\end{equation}
thus during the inflationary era the Legendre polynomial related
term merely contributes a cosmological constant. Also at late
times, the same term in the small curvature regime at leading
order contributes a cosmological constant, since,
\begin{equation}\label{legendrelatetime}
\lim_{R\to 0}(R+R_0)P_3\left(\frac{M_P^2}{R+R_0}\right)\simeq
\frac{15 M_P^6 R^2}{2 R_0^4}-\frac{10 M_P^6 R^3}{R_0^5}-\frac{5
M_P^6 R}{R_0^3}+\frac{5 M_P^6-3 M_P^2 R_0^2}{2 R_0^2}\, .
\end{equation}
Thus during inflation, the term that drives the inflationary era
is the $R^2$ term and at late times the cosmological constant
appearing in Eq. (\ref{legendrelatetime}) basically determines the
evolution. As we will show numerically, the dark energy era
corresponds to a pure cosmological constant non-dynamical
behavior, in contrast to the $f(R)$ gravity case considered in the
previous section.

In order to properly study the inflationary era we shall focus
mainly on the slow-roll indices \cite{Hwang:2005hb}, designated
as,
\begin{align}
\centering
\label{slowroll}
\epsilon_1&=-\frac{\dot H}{H^2}&\epsilon_2&=\frac{\ddot\phi}{H\dot\phi}&\epsilon_3&=\frac{\dot f_R}{2Hf_R}&\epsilon_4&=\frac{\dot E}{2HE}&\epsilon_6&=\sum_{l}\frac{\dot Q_t}{2HQ_t}\, ,
\end{align}
where $f_R=\frac{d f}{dR}$,
$E=-\frac{f_R}{X}\left(XG_{,X}+2X^2G_{,XX}+\frac{3\dot
f_R^2}{2\kappa^2f_R} \right)$ and
$Q_t=f_R+2\lambda_l\kappa^2\dot\nu H$, where $l$ sums over the
polarizations of the gravitational waves with $\lambda_l=\pm1$ for
left and right handed polarization respectively. It is worth
mentioning that the second slow-roll index is identically equal to
zero since the scalar field evolves primordially linearly with
respect to time. Knowledge of the slow-roll indices is essentially
important as the observational indices can be extracted from them.
Let us now study each slow-roll index separately and derive
certain conclusions.

Obviously, as mentioned before, $\ddot\phi=0$ therefore
$\epsilon_2=0$. Now according to Eq.(\ref{slowroll}), we have,
\begin{equation}
\centering
\label{index1a}
\epsilon_1=-\frac{\dot H}{H^2}\, ,
\end{equation}
and since the $R^2$ term drives the early-time era and
specifically the Friedmann equation, the Hubble rate during
inflation is perfectly described by the quasi-de Sitter expansion
$H(t)=H_0-H_it$ (see Ref. \cite{Odintsov:2020nwm} for details). We
shall further support this result of a quasi-de Sitter evolution
later on, when we consider specific forms of the Chern-Simons
coupling function (see the arguments below Eq. (\ref{expCS})).
Thus, the first slow-roll index reads,
\begin{equation}
\centering
\label{index1b}
\epsilon_1=\frac{H_i}{H_0^2\left(1-\frac{H_i t_i}{H_0}\right)^2}\, ,
\end{equation}
In order to ascertain the time duration of the inflationary era we
assume that the first slow-roll index reaches unity. Solving
$\epsilon_1(t_f)\sim\mathcal{O}(1)$ we find that,
\begin{equation}
\centering
\label{tf}
t_f=\left(1\mp\frac{\sqrt{H_i}}{H_0}\right)\frac{H_0}{H_i}\, ,
\end{equation}
This signifies the time instance in units of eV$^{-1}$ when the
inflationary era ceases and of course the positive solution is
relevant. In consequence, from the definition of the $e$-foldings
number $N=\int_{t_i}^{t_f}{H(t)dt}$, the initial time $t_i$ for
which inflation starts, can be computed and reads,
\begin{figure}[h!]
\centering
\includegraphics[width=25pc]{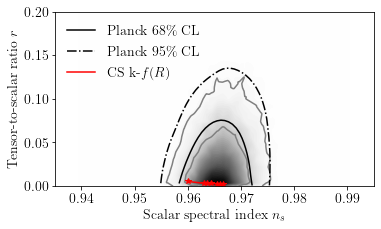}
\caption{Planck 2018 likelihood curves and the Chern-Simons
$k$-essence $f(R)$ gravity inflationary predictions for
$N=[50,60]$ and $f_1=2\times 10^{N}$.} \label{planclikelihood}
\end{figure}
\begin{equation}
\centering
\label{ti}
t_i=c_1+\sqrt{c_1^2+c_2}\, ,
\end{equation}
where we keep only the positive solution as $c_2>0$ and the
auxiliary parameters read $c_1=\frac{H_0}{H_i}$ and
$c_2=t_f^2+\frac{2N}{H_i}-\frac{2H_0t_f}{H_i}$. The initial time
is of paramount importance as the observational indices are
extracted by evaluating the slow-roll indices an mentioned before,
at exactly that time instance, which signifies the first horizon
crossing during inflation. Let us now proceed with the rest
slow-roll indices. For the slow-roll index $\epsilon_3$, we have,
\begin{align}
\centering
\label{auxiliary2}
R&\simeq12H^2&f_R&\simeq1+2\alpha R&\dot f_R&\simeq2\alpha \dot R\, ,
\end{align}
so if we assume contribution from the $R^2$ part since it is more
dominant during the inflationary era, we find that
$\epsilon_3\simeq-\epsilon_1$ and their difference due to unity in
$f_R$ is of order $\mathcal{O}\left(\frac{1}{N^2}\right)$
something which has an impact on the tensor-to-scalar ratio. Now
for slow-roll index $\epsilon_4$, we have $\epsilon_4=\frac{\dot
E}{2HE}$ with,
\begin{equation}
\centering
\label{E}
E=-\frac{f_R}{X}\left(XG_{,X}+2X^2G_{,XX}+\frac{3\dot f_R^2}{2\kappa^2f_R}\right)=f_R\left(1+6f_1\kappa^4X-\frac{3\dot f_R^2}{2\kappa^2f_RX}\right)\, ,
\end{equation}
The final part scales with $\frac{1}{N^2}$ so it is significantly
subdominant, therefore we find that $\epsilon_4\simeq-\epsilon_1$.
Overall the scalar spectral index reads,
\begin{equation}
\centering
\label{ns}
n_S=1-2(2\epsilon_1+\epsilon_2-\epsilon_3+\epsilon_4)\simeq1-\frac{2}{N}+\mathcal{O}\left(\frac{1}{N^2}\right)\, ,
\end{equation}
Note that $\frac{1}{N^2}$ corrections come from the difference
between $\epsilon_3$ and $\epsilon_1$ while
$\mathcal{O}(\frac{1}{N^3})$ comes from the neglected term in
$\epsilon_4$ so the overall phenomenology seems quite right. As
shown, it coincides with the exact form of the $R^2$ model. Now
for the tensor-to-scalar ratio, according to Ref.
\cite{Hwang:2005hb}, we have,
\begin{equation}
\centering \label{ra}
r=8|\epsilon_1+\epsilon_3|c_A\sum_{l}\frac{1}{\left|1+\frac{2\lambda_l\kappa^2\dot\nu
H}{f_R}\right|}\, .
\end{equation}
From the definition of the sound wave velocity, we find that,
\begin{equation}
\centering
\label{cA}
c_A=\sqrt{\frac{XG_{,X}+\frac{3\dot f_R^2}{2\kappa^2f_R}}{XG_{,X}+2X^2G_{,XX}+\frac{3\dot f_R^2}{2\kappa^2f_R}}}=\sqrt{\frac{1-f_1^{\frac{1}{3}}}{1-3f_1^{\frac{1}{3}}}}+\mathcal{O}\left(\frac{1}{N^2}\right)\, ,
\end{equation}
if we perform a Taylor expansion for large $N$ in the $\frac{\dot
f_R^2}{2\kappa^2f_RX}$ term. Moreover, the Chern-Simons
contribution during the first horizon crossing, if one replaces
$H(t_i)$ in $f_R$ then the following expression emerges,
\begin{equation}
\centering
\label{rb}
r\simeq\frac{6}{N^2}\sqrt{\frac{1-f_1^{\frac{1}{3}}}{1-3f_1^{\frac{1}{3}}}}\left(\frac{1}{\left|1+\frac{3\kappa^2\dot\nu\sqrt{H_i}}{\sqrt{2N}}\right|}+\frac{1}{\left|1-\frac{3\kappa^2\dot\nu\sqrt{H_i}}{\sqrt{2N}}\right|}\right)\, ,
\end{equation}
Here, two main features are mentionable. Firstly, the case of the
$R^2$ model's tensor-to-scalar ratio, meaning $r\simeq
\frac{12}{N^2}$ can be extracted from the above expression by
choosing $\nu=0$ and $f_1=0$, therefore the known results of the
$R^2$ gravity are safely extracted. Furthermore, due to the
presence of a non-canonical kinetic term, the field propagation
velocity satisfies the relation $c_A<1$ thus the $k$-essence term
results in a decrease of the tensor-to-scalar ratio relative to
the expected $\frac{12}{N^2}$. For the time being, no conclusion
can be extracted for the time derivative of the Chern-Simons
scalar coupling function and therefore the additional contribution
in Eq. (\ref{rb}). The result should be considered to be
model-dependent.

Finally, for the tensor spectral index, we need
$\epsilon_1\simeq\frac{1}{2N}$ and $\epsilon_6=\sum_{l}\frac{\dot
Q_t}{2HQ_t}$ where $Q_t(t_i)=f_R(t_i)+2\lambda_l\kappa^2\dot\nu
H(t_i)$. If we evaluate the ratio $\frac{\dot Q_t}{2HQ_t}$ we
find,
\begin{equation}
\centering \label{ratio} \frac{\dot
Q_t}{2HQ_t}\Big{|}_{t_i}=\epsilon_3\left(\frac{1+\frac{\lambda_l\kappa^2\dot\nu
H}{f_R}}{1+\frac{2\lambda_l\kappa^2\dot\nu
H}{f_R}}\right)+\frac{\lambda_l\kappa^2\ddot\nu}{f_R\left(1+\frac{2\lambda_l\kappa^2\dot\nu
H}{f_R}\right)}\, ,
\end{equation}
where hereafter everything is evaluated at the first horizon
crossing. It is tempting,  to say the least, to argue that under
the slow-roll assumption the first part is almost equal to
$\epsilon_3$, however this is not the case here since the
functions $\ddot\nu$ and $\dot\nu$ could be more dominant,
therefore one needs to keep such term as it is and evaluate it
numerically. In general, we have,
\begin{equation}
\centering
\label{index6a}
\epsilon_6=\epsilon_3\sum_{l}\left(\frac{1+\frac{\lambda_l\kappa^2\dot\nu H}{f_R}}{1+\frac{2\lambda_l\kappa^2\dot\nu H}{f_R}}\right)-\frac{\frac{4\kappa^4\ddot\nu\dot\nu H}{f_R^2}}{1-\frac{4\kappa^4\dot\nu^2H^2}{f_R^2}}\, ,
\end{equation}
and by focusing on the case of $1\ll
\frac{\lambda_l\kappa^2\dot\nu H}{f_R}$, one finds that,
\begin{equation}
\centering
\label{index6b}
\epsilon_6=\epsilon_3-\frac{\frac{4\kappa^4\ddot\nu\dot\nu H}{f_R^2}}{1-\frac{4\kappa^4\dot\nu^2H^2}{f_R^2}}\, ,
\end{equation}
One would make the approximation $\epsilon_3\simeq-\epsilon_1$
however this could result in a zero tensor spectral index for the
case of $\ddot\nu\ll\dot\nu^2$, and this is true at first order.
However, if we recall that for inflationary $f(R)$ attractors, as
we saw previously, the slow-roll indices are interconnected as
$\epsilon_1=-\epsilon_3(1-\epsilon_4)$ therefore in our case,
$\epsilon_3=-\epsilon_1+\epsilon_3\epsilon_4$ and
$\epsilon_4\simeq-\epsilon_1$, therefore to leading order,
$\epsilon_3=-\epsilon_1+\epsilon_1^2$ thus the tensor spectral
index reads,
\begin{equation}
\centering
\label{nta}
n_T=-2\epsilon_1^2+\frac{\frac{8\kappa^4\ddot\nu\dot\nu H}{f_R^2}}{1-\frac{4\kappa^4\dot\nu^2H^2}{f_R^2}}\, ,
\end{equation}
where it becomes apparent that for the case of $\nu=0$, the result
coincides with the one obtained for the $R^2$ model developed in
the previous section. Before we proceed, it is worth mentioning
that due to the fact that $\epsilon_6$ is proportional to
$-\epsilon_1$, the tensor spectral index receives positive
contribution to order $\mathcal{O}\left(\frac{1}{N^2}\right)$.

If we focus on the numerator and denominator separately and
evaluate everything during the first horizon crossing, then a term
$\frac{9\kappa^4\ddot\nu\dot\nu\sqrt{H_i}}{\sqrt{2}N^{\frac{3}{2}}}\frac{1}{1-\frac{9\kappa^4\dot\nu^2H_i}{f_R^2}}$
appears, if we expand in powers of $N$ in the large $N$ limit,
then a term of order
$\mathcal{O}\left(\frac{1}{N^{\frac{5}{2}}}\right)$ emerges in the
tensor spectral index, and we treat it as subleading. Overall we
find,
\begin{equation}
\centering
\label{ntb}
n_T\simeq-\frac{1}{2N^2}+\frac{9\kappa^4\ddot\nu\dot\nu\sqrt{H_i}}{\sqrt{2}N^{\frac{3}{2}}}+\mathcal{O}\left(\frac{1}{N^{\frac{5}{2}}}\right)\, ,
\end{equation}
Depending on the sign/dominance of the first term, the tensor
spectral index may be blue-tilted. Note also that in principle
$\ddot\nu=\dot\phi^2\nu''$ given that $\ddot\phi=0$. Now, if we
use the fact that $H_i=\frac{M^2}{6}$, such that
$f(R)=R+\frac{R^2}{6M^2}$ we find that with respect to such mass
scale the observational indices become,
\begin{align}
\centering
\label{observables}
n_S&\simeq1-\frac{2}{N}+\mathcal{O}\left(\frac{1}{N^2}\right)\\
r&\simeq\frac{6}{N^2}\sqrt{\frac{1-f_1^{\frac{1}{3}}}{1-3f_1^{\frac{1}{3}}}}\left(\frac{1}{\left|1+\frac{3\kappa^2\dot\nu M}{2\sqrt{3N}}\right|}+\frac{1}{\left|1-\frac{3\kappa^2\dot\nu M}{2\sqrt{3N}}\right|}\right)+\mathcal{O}\left(\frac{1}{N^4}\right)\\
n_T&\simeq-\frac{1}{2N^2}+\frac{9\kappa^4\ddot\nu\dot\nu M}{2\sqrt{3}N^{\frac{3}{2}}}+\mathcal{O}\left(\frac{1}{N^{\frac{5}{2}}}\right)\, ,
\end{align}
Truthfully, the final form  of the tensor spectral index is
possible only if $\kappa^3\ddot\nu\dot\nu$ is well behaved,
however in this case the ratio in front of $\epsilon_3$ in
(\ref{index6a}) would be closer to unity for a given polarization
$l$, thus instead of $\frac{1}{2N^2}$ a factor of $\frac{1}{N}$
would emerge. Therefore only in such case the expansion is
permitted, however for the $k$-essence model where the $\dot\nu$
function is quite large, as we shall showcase below, Eq.
(\ref{nta}) should be used in the following form
\begin{equation}
\centering
\label{ntc}
n_T\simeq-\frac{1}{2N^2}+\frac{\frac{9\kappa^4\ddot\nu\dot\nu M}{2\sqrt{3N}N}}{1-\frac{3\dot\nu^2\kappa^2M^2}{4N}}\, ,
\end{equation}
The resulting behavior is model-dependent as we shall showcase
briefly. To prove this, we shall use certain power-law and
exponential models for the Chern-Simons scalar coupling function,
and the exponential shall be used for the primordial and late-time
evolution study, and for the calculation of the primordial
gravitational wave energy power spectrum.

Let us start with the case of,
\begin{equation}
\centering
\label{case1}
\nu(\phi)=\Lambda_1(\kappa\phi)^2\, ,
\end{equation}
then for the case of $\Lambda_1=100$, $f_1=-20$ and $N=60$ we find
that $r=0.000209$ and $n_T=-0.083879$ while for an increasing
$\Lambda_1$ the tensor to scalar ratio decreases due to the
Chern-Simons term. The result is extracted by making use of Eq.
(\ref{ntc}) otherwise one would find a diverging tensor spectral
index given that $\dot\phi\sim 10^{55}$eV$^2$ during the first
horizon crossing. So we see that the expansion in large $N$ in the
presence of a $k$-Essence term is not a viable option for the case
of the tensor spectral index, due to the Chern-Simons term. In
fact, the Chern-Simons contribution becomes so dominant that the
sign of the tensor spectral index switches and becomes negative.

Let us examine now a different model. Suppose that,
\begin{equation}
\centering
\label{case2}
\nu(\phi)=\Lambda_1\phi\, ,
\end{equation}
with $\Lambda_1\sim\mathcal{O}(10^{400})$, $f_1\sim\mathcal{O}(1)$
and $N\sim60$, then due to the fact that $\ddot\phi$ is
identically zero the tensor spectral index is positive and equal
to $n_T=0.00014$ whereas due to the extreme value of $\Lambda_1$
the tensor-to-scalar ratio is suppressed to $r=0$. The
aforementioned model is indicative of the fact that the scalar
field assisted $f(R)$ gravity can be used in order to properly
describe both early and late-time eras in a unified manner and
also make significant predictions for gravitational waves and
future experiments.

\begin{figure}[h!]
\centering
\includegraphics[width=20pc]{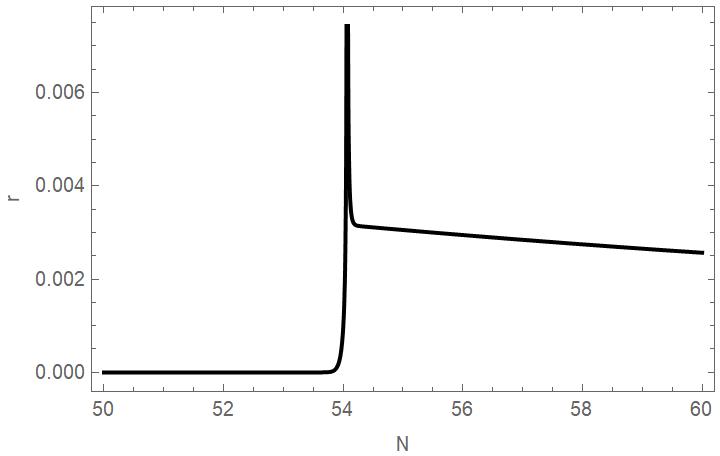}
\includegraphics[width=20pc]{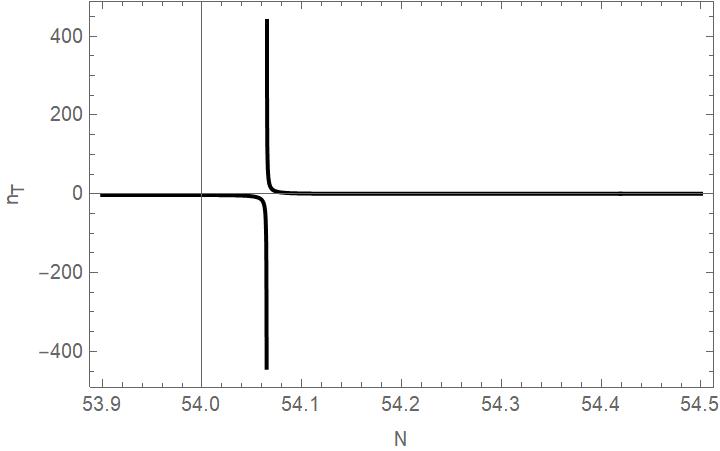}
\caption{Tensor to scalar ratio $r$ (left) and tensor spectral
index $n_T$ (right) depending on the e-folding number and
parameter $f_1$ satisfying the condition $f_1=2\cdot10^N$.}
\label{constrainedindices}
\end{figure}

Finally we introduce the model we shall work with in the following
sections. This particular model admits an exponential Chern-Simons
scalar coupling function of the form,
\begin{equation}
\centering
\label{expCS}
\nu(\phi)=\Lambda_1 e^{\sqrt{\frac{\phi}{M_1}}}\, ,
\end{equation}
where $\Lambda_1=4\cdot10^{-12}$ and
$M_1=25\cdot10^{-5.332}\Lambda_1 M_P$. For this particular model
and with the assumption that $f_1=2\cdot10^{N}$ and $N=54.2$ one
finds using equations (\ref{rb}) and (\ref{ntc}) that the tensor
to scalar ratio and the tensor spectral index respectively obtain
the values $r=0.003165$ and $n_T=0.01659$. We shall use these
values for the calculation of the gravitational wave energy
spectrum later on. In Fig. \ref{planclikelihood} we present the
Planck 2018 likelihood curves and the inflationary predictions of
the Chern-Simons $k$-essence $f(R)$ gravity model for $N=[50,60]$
and $f_1=2\times 10^{N}$ (red curve). As it can be seen in Fig.
\ref{planclikelihood} the Chern-Simons $k$-essence $f(R)$ gravity
model is well fitted in the Planck data. It is worth mentioning
that the number of free parameters is effectively decreased to 3,
namely $\Lambda_1$, $M$ and $N$ given that $M_1$ and $f_1$ are now
functions of $\Lambda_1$ and $N$ respectively whereas the $R^2$
scale is assumed to be equal to $M=1.25\cdot10^{-5}M_P$ similar to
the previous power-law cases. Note that such designation suggests
that the tensor spectral index is blue tilted which in turn
suggests that the denominator in equation (\ref{ntc}) is more
dominant than the numerator. The case of large $f_1$ and $\dot\nu$
results in a finite ratio between the field propagation velocity
and the Chern-Simons contribution in the tensor-to-scalar ratio
(\ref{rb}), therefore it is non zero, but as expected it satisfies
the relation $r<\frac{12}{N^2}$. Note also that decreasing $f_1$
and essentially not-constraining it, the $e$-foldings number can
result in principle in a negative tensor spectral index. As a
final note it should be stated that for the case of $N>54$,
working with the expression for $n_T$ extracted in Eq.
(\ref{observables}) is a viable choice as it produces the same
result with (\ref{ntc}), something which can be seen from Fig.
\ref{constrainedindices}, where we plot the tensor-to-scalar ratio
$r$ (left) and tensor spectral index $n_T$ (right) as functions of
the $e$-foldings number and with the parameter $f_1$ being chosen
$f_1=2\cdot10^N$. As shown, the value of $N=54$ serves as a
boundary for which the sign of the tensor spectral index changes
and above which the approximated solution for $n_T$ extracted
before can safely be used. This is a model dependent result and
applies only for the exponential Chern-Simons scalar coupling
function. For the order of magnitude, $\dot\phi\sim10^{37}eV^2$
and $R^2\sim 10^{90}eV^2$ assuming that $H\sim 10^{13}GeV$
therefore, the $R^2$ contribution is more dominant in the
inflationary era compared to the $k$-essence part, thus validating
the choice of the quasi-de Sitter expansion.
The exponential model we just discussed, shall be used
subsequently to study the late-time phenomenology of the
Chern-Simons $k$-essence $f(R)$ model, and for the evaluation of
the primordial gravitational wave energy spectrum. Due to the fact
that the signal of pure $f(R)$ gravity models is quite suppressed
relative to the current detectable amplitudes from present and
future missions and experiments, such as the NANOGrav, LISA,
DECIGO etc., it can be shown that the inclusion of a cosmological
scalar field can actually contribute both in the late-time
description and can significantly affect the energy spectrum of
the gravitational waves of inflation.


\subsection{From the Dark Energy Era to Radiation Domination Era for the Chern-Simons $k$-essence $f(R)$ Gravity and the Primordial Gravitational Waves Spectrum}

\begin{figure}[h!]
\centering
\includegraphics[width=20pc]{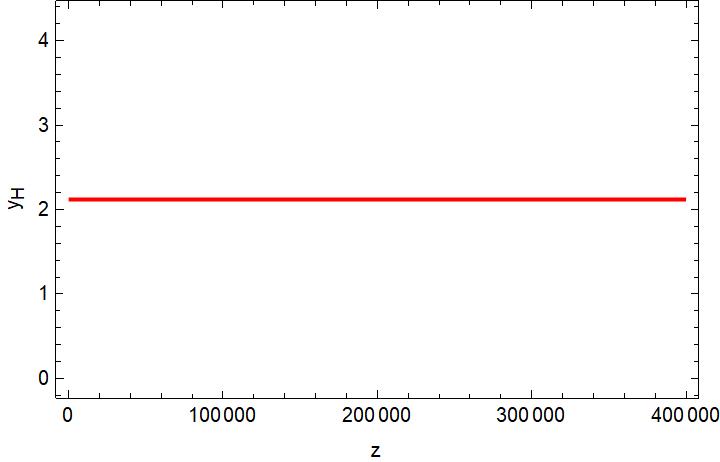}
\includegraphics[width=20pc]{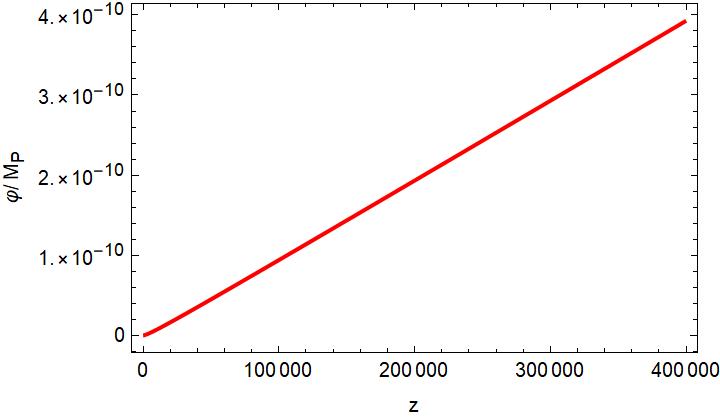}
\caption{Numerical solutions for statefinder $y_H$ and the scalar
field normalized with the Planck mass as functions of redshift. It
becomes apparent that while $y_H$ is constant in time, meaning
that it behaves effectively as a cosmological constant
corresponding to constant dark energy density and in consequence
constant dark energy EoS, the scalar field seems to decrease with
time.} \label{sol}
\end{figure}
Having discussed the inflationary phenomenology of the model
(\ref{latetimef2initial}), let us now consider the evolution of
the model during the various cosmological eras, from the dark
energy era back to the radiation domination era. We shall
transform all the field equations in such a way so that the
variable will be the statefinder $y_H(z)$ we used in the pure
$f(R)$ gravity section, and the dynamical variable will be the
redshift $z$. Our aim is to calculate the overall ``damping''
factor $\mathcal{D}$ thus to calculate the integral of the
parameter (\ref{amchersimonsfinal}). The upper limit of the
redshift integration will be determined approximately to be the
redshift for which the cosmological model enters the radiation
domination era, and as it proves it is approximately
$z_{fin}=400000$. Beyond that, parameter $a_{M\ell}$ gets
simplified to the form (\ref{amchersimonsfinalsimpl}) and as it
proves, the integral (\ref{dform}) receives no contribution beyond
the upper redshift $z_{fin}=400000$.

The main focus in this section lies in the numerical solution of
the equations of motion and in particular both the Friedmann
equation and the continuity equation of the scalar field which was
introduced previously during the inflationary era. For the sake of
generality, an arbitrary minimally coupled scalar-tensor model, an
in particular a $k$-essence $f(R)$ model shall be presented but
subsequently we shall focus mainly on the $R^2$ model with an
exponential Chern-Simons coupling.

Let us commence from the equations of motion. For an arbitrary
minimal scalar-tensor model in the presence of perfect fluids, the
Friedman and continuity equation for the scalar field read
\begin{equation}
\centering
\label{Friedmann}
\frac{3f_RH^2}{\kappa^2}=\rho+\frac{f_RR-f}{2\kappa^2}-\frac{3H\dot f_R}{\kappa^2}+G_{,X}X-\frac{G}{2}\, ,
\end{equation}
\begin{equation}
\centering
\label{Conteq}
G_{,X}(\ddot\phi+3H\dot\phi)+\dot\phi\dot G_{,X}=0\, ,
\end{equation}
\begin{figure}[h!]
\centering
\includegraphics[width=25pc]{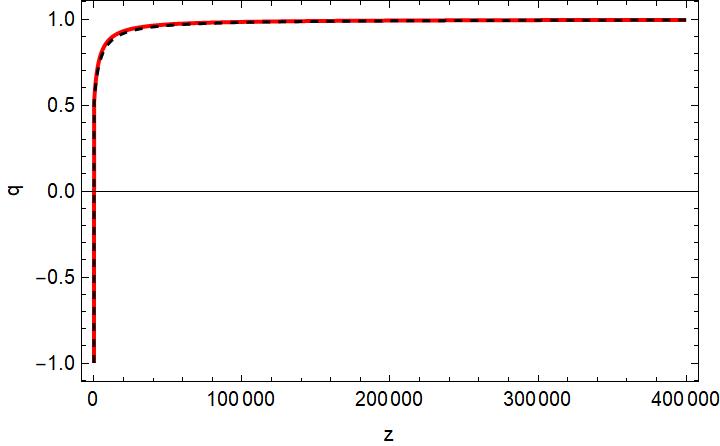}
\caption{The deceleration parameter versus the redshift for the
Chern-Simons $k$-essence $f(R)$ gravity model versus the redshift
(red curve). The black dashed curve corresponds to the
$\Lambda$CDM model} \label{statefinders}
\end{figure}
The main focus lies with the numerical solution of Hubble's
parameter and the scalar field, as functions of cosmic time $t$.
This procedure is intricate and thus certain transformations are
needed in order to facilitate the study of the late-time era and
essentially work backwards and study the matter and radiation
dominated eras respectively. In order to proceed, two
transformations need to be performed that facilitate the overall
procedure. The first transformation is a variable transformation
and is used in order to replace cosmic time $t$ with redshift
though the definition of redshift,
\begin{equation}
\centering
\label{z}
1+z=\frac{1}{a}\, ,
\end{equation}
where for simplicity it was assumed that the current value of the
scale factor is normalized to unity in order for the comoving
wavelengths and physical wavelengths to coincide at present time.
Essentially one can think of such definition as a function of the
form $t=t(z)$, therefore a connection between time derivatives and
derivatives with respect to redshift is needed. This connection
emerges naturally from Eq. (\ref{z}) by means of construction of a
differential operator that connects these two variables. This
operator reads,
\begin{equation}
\centering
\label{dz}
\frac{d}{dt}=-H(1+z)\frac{d}{dz}\, ,
\end{equation}
with the Hubble rate now being a function of redshift itself. For
simplicity, differentiation with respect to redshift shall be
denoted with prime $\frac{d}{dz}='$. Therefore, certain time
derivatives which participate in the aforementioned equations of
motion are transformed as,
\begin{align}
\centering
\label{derivatives}
\dot H&=-H(1+z)H'\\
\dot\phi&=-H(1+z)\phi'\\
\dot R&=-H(1+z)R'
\end{align}
where the last participates in the Friedmann equation through
$\dot f_R=f_{RR} \dot R$. Performing such replacements in
equations (\ref{Friedmann})-(\ref{Conteq}) eliminates cosmic time
$t$, as expected, however working with Hubble itself is a bit
intricate as it has mass dimensions $[m]=$eV.
\begin{figure}[h!]
\centering
\includegraphics[width=25pc]{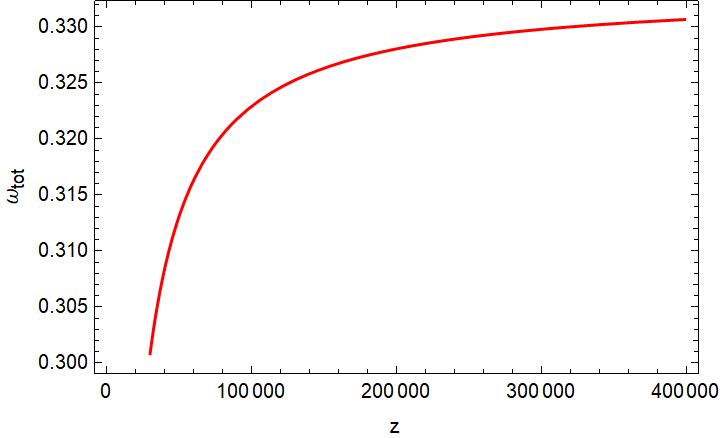}
\caption{The total EoS parameter for the Chern-Simons
S-$k$-essence $f(R)$ gravity model.} \label{omegatotcs}
\end{figure}
Instead, one can work with a new dimensionless variable which can
be derived from the Friedmann equation. In particular, by
rewriting Eq. (\ref{Friedmann}) in the usual form the Friedmann
equation has, the following expression is derived,
\begin{equation}
\centering
\label{Friedmann2}
\frac{3H^2}{\kappa^2}=\rho+\rho_{DE}\, ,
\end{equation}
where the new contribution specifies dark energy at late times and
is specified as,
\begin{equation}
\centering
\label{Dedensity}
\rho_{DE}=\frac{f_RR-f}{2\kappa^2}-\frac{3H\dot f_R}{\kappa^2}+G_{,X}X-\frac{G}{2}+\frac{3H^2}{\kappa^2}(1-f_R)\, ,
\end{equation}
Here, it becomes abundantly clear that the dark energy density,
and in general the definition of the dark energy, is nothing but
the contribution of all geometric terms which appear in the
gravitational action and in consequence the Friedmann equation.
Subsequently, we define the dimensionless statefinder parameter
$y_H$, as in the pure $f(R)$ gravity section,
\begin{equation}
\centering \label{yh}
y_H(z)=\frac{\rho_{DE}}{\rho_0}=\frac{H^2}{m_s^2}-\frac{\rho}{\rho_{m}^{(0)}}\,
,
\end{equation}
where $\rho_{m}^{(0)}$ is the current density of non relativistic
matter and $m_s^2=\frac{\kappa^2\rho_{m}^{(0)}}{3}$ serves as a
mass scale which were defined in the pure $f(R)$ gravity section.
This statefinder in particular is used for replacing Hubble's
parameter in the equations of motion which shall be solved
numerically. Furthermore the matter density is given by the
formula,
\begin{equation}
\centering \label{rho} \rho=\rho_{m}^{(0)}(1+z)^3(1+\chi(1+z))\, ,
\end{equation}
where $\chi=\frac{\rho_{r0}}{\rho_{m}^{(0)}}$ serves as a ratio
between the current value of relativistic and non relativistic
matter density. Therefore, by replacing Hubble's parameter with
the statefinder $y_H$, the numerical solution of equations
(\ref{Friedmann})-(\ref{Conteq}) becomes easier to extract. In
particular, we mention that the replacement of Hubble's parameter
is achieved though the subsequent replacements,
\begin{align}
\centering
\label{H}
H&=m_s\sqrt{y_H+\tilde\rho}\\
HH'&=\frac{m_s^2}{2}(y_H'+\tilde\rho')\\
H'^2+HH''&=\frac{m_s^2}{2}(y_H''+\tilde\rho'')
\end{align}
where for simplicity $\tilde\rho=\frac{\rho}{\rho_{m}^{(0)}}$.
With these equations at hand, one can replace both cosmic time $t$
and Hubble with redshift and statefinder $y_H$ respectively. The
late-time analysis admits numerical solutions for variables $y_H$
and $\phi$ with respect to redshift in the interval [0,400000].
Motivated by the behavior of the $f(R)$ gravity
(\ref{latetimef2initial}) at both the inflationary era and the
dark energy era, which behaves as a cosmological constant, we
shall use the following initial conditions for the function
$y_H(z)$,
\begin{equation}\label{generalinitialconditionsfinalstudy}
y_H(0)=\frac{\Lambda}{3m_s^2}\, , \,\,\,\frac{d y_H(z)}{d
z}\Big{|}_{z=0}=0\, .
\end{equation}
Also for the scalar field we shall use the following initial
conditions,
\begin{equation}\label{initialconditionsscalarfield}
\phi(0)=\frac{M_P}{10^{16}},\,\,\,\phi'(0)=\frac{M_P}{10^{17}}\, ,
\end{equation}
which are also motivated physically (see Eq. (16) of Ref.
\cite{Arai:2017hxj}).
\begin{figure}[h!]
\centering
\includegraphics[width=25pc]{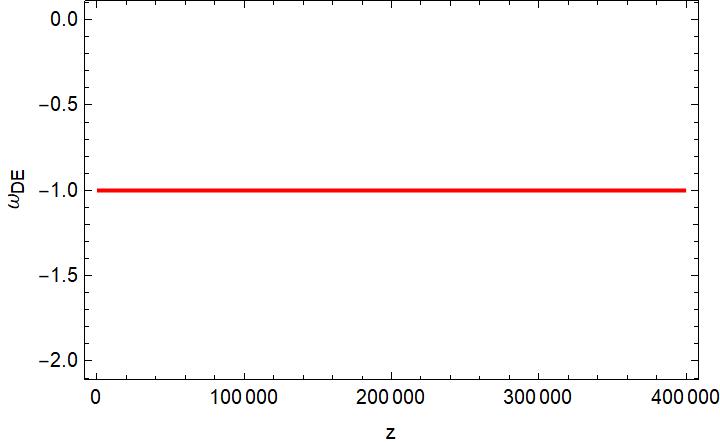}
\caption{Dark energy EoS depending solely on redshift in the
interval $z=[0,4\times 10^5]$. It becomes apparent that while the
first remains constant as a direct consequence of the constant
value if statefinder $y_H$, the latter remains infinitesimally
close to zero and starts increasing for relatively small
redshifts. } \label{DE}
\end{figure}
In subsequent models the numerical solutions will cover redshifts
starting from present day at $z=0$ until the early radiation
dominated era thus in the interval $[0,4\times 10^5]$. However
before we proceed it is important to make a statement on the
dimensionality of the initial condition of the derivative of the
scalar field. Previously, it was given in terms of the Planck
mass, which is indeed the case as function $\phi'$ runs with eV.
Note that the time derivative of the scalar field must have mass
dimensions of eV$^2$ however due to the differential operator
(\ref{dz}) which already runs with eV units, $\phi'$ must also
have the same exact units. As a final note, it should be stated
that for the model at hand, it was assumed that perfect matter
fluids were present therefore from Eq. (\ref{Friedmann2}) the dark
energy density should also satisfy a perfect fluid continuity
equation of the form,
\begin{equation}
\centering
\label{conteq}
\dot\rho_i+3H\rho_i(1+\omega_i)=0\, ,
\end{equation}
where index $i$ takes the values $m$, $r$ and $DE$ for non
relativistic, relativistic and dark energy respectively. It should
be stated that parameter $\omega_i$ characterizes the equation of
state of each fluid, where for non relativistic matter
$\omega_{m}=0$, for relativistic matter $\omega_r=\frac{1}{3}$. In
the following, the EoS for dark energy shall be properly specified
as a statefinder parameter and afterwards we shall ascertain
whether for a given model compatible with Planck data
\cite{Aghanim:2018eyx}, it is feasible to achieve a value of
$\omega_{DE}\simeq-1$ at present day.
\begin{figure}[h!]
\centering
\includegraphics[width=20pc]{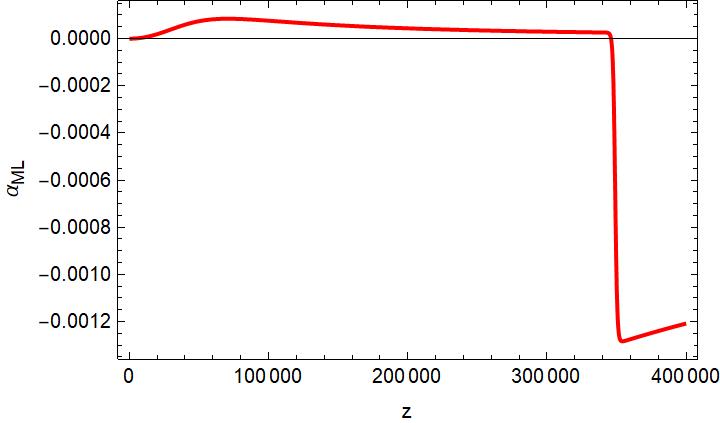}
\includegraphics[width=20pc]{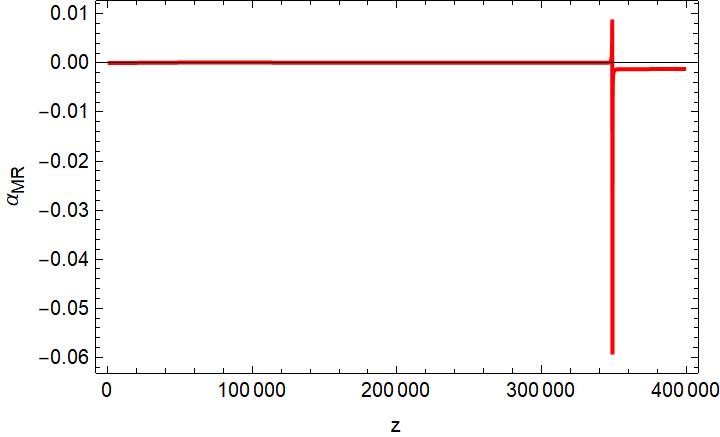}
\caption{The behavior of the parameters $a_{ML}$ (left plot) and
of $a_{MR}$ (right plot) for $z=[0,4\times 10^{5}]$. }
\label{amLRupdatecs}
\end{figure}
In order properly study the late-time era and in general the
validity of a particular model with the Planck data, and in
particular the $\Lambda$CDM model, currently the best description
of the Universe, certain statefinder parameters shall be extracted
and in particular compared to such observable quantities. In the
present article we shall limit our work to one statefinder
parameter that specifies the evolution of the Universe and two
that are solely connected to dark energy. In particular, we shall
consider the deceleration parameter $q$, with regard to
statefinders, and the dark energy EoS parameter $\omega_{DE}$, and
the total EoS parameter $\omega_{tot}$. A proper description of
the Universe admits a total EoS $\omega_{tot}$, which tends to
$\omega_{tot}=-1$ for $z=0$, and tends asymptotically to
$\omega_{tot}=\frac{1}{3}$ at approximately $z_{fin}=4\times
10^5$. Regarding dark energy, two statefinder parameters shall be
studied, in particular the equation of state and the dark energy
density parameter $\omega_{DE}$ and $\Omega_{DE}$ respectively
which recall are given by the following expressions,
\begin{align}
\centering
\label{de}
\omega_{DE}&=-1+\frac{1+z}{3}\frac{d\ln y_H}{dz}&\Omega_{DE}&=\frac{y_H}{y_H+\tilde\rho}\, ,
\end{align}
In the following subsection we shall study the cosmological
behavior of the model (\ref{latetimef2initial}) from the dark
energy era back to the early stages of the radiation domination
era, and in some cases we shall compare the model with the
$\Lambda$CDM model.
\begin{figure}[h!]
\centering
\includegraphics[width=40pc]{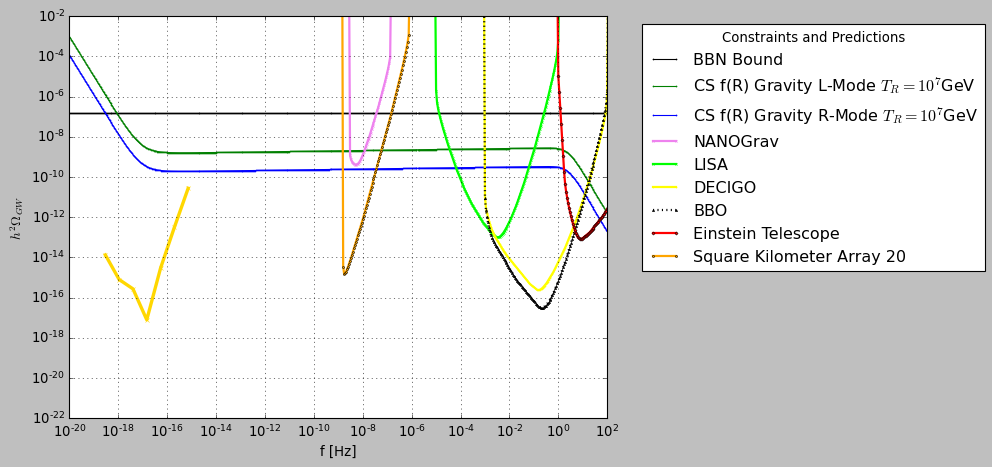}
\caption{The $h^2$-scaled gravitational wave energy spectrum for
the Chern-Simons corrected $k$-essence $f(R)$ gravity for the
reheating temperature being $T_R=10^7$GeV.}
\label{plotfinalfrcstr7}
\end{figure}

\subsection{Numerical Results}

Let us quote here for reading convenience the analytic form of the
model (\ref{latetimef2initial}), which is,
\begin{equation}
\centering \label{latetimef2}
f(R)=R+\frac{R^2}{6M^2}-(R+R_0)P_3\left(\frac{M_P^2}{R+R_0}\right)\,
,
\end{equation}
where by demanding compatibility with the Planck data, auxiliary
parameters were specified as $M=1.25\cdot 10^{-5}\,M_P$,
$\Lambda_1=4\cdot10^{-12}$, $M_1=25\cdot
10^{-5.332}\Lambda_1\,M_P$. Also the $R_0$ serves as an effective
mass scale which is assumed to be of order
$R_0\sim10^{-52}$eV$^2$. Finally the non-canonical parameter which
participates in the $k$-essence contribution will be assumed to be
related with the $e$-foldings number in the following way
$f_1=2e^N$ with $N$ being $N=54.2$. Therefore, with these
functions at hand and the numerical values of the auxiliary
parameters, the equations of motion can be numerically solved and
extract information about statefinder $y_H$ and the scalar field
$\phi$ with respect to redshift. Now that such detail is out of
the way, let us proceed directly to the numerical results of the
aforementioned model.

We commence with the statefinders introduced previously that
describe both the evolution of the Universe and the behavior of
dark energy, which in this framework is identified as all the
extra geometric terms participating in (\ref{action}), coming both
from the $f(R)$ and the $G(X)$ functions. Initially, for the
solutions of equations (\ref{Friedmann}) and (\ref{Conteq})
depicted in Fig \ref{sol}, we find that the statefinder $y_H$ is
stationary, as it was expected since the quantity $y_H$ is
basically the dark energy density which in the case at hand
behaves as a pure cosmological constant for the whole evolution of
the Universe. Our numerical study presented in the left plot of
Fig. \ref{sol} indicates this fact directly. In the right plot of
Fig. \ref{sol} we present the behavior of the normalized scalar
field as a function of the redshift, which decreases as the
redshift approaches the present day value $z=0$. The fact that the
statefinder $y_H$ as a solution is stable suggests that the
evolution of Hubble's parameter is solely specified by the
component $\rho(z)$ but is shifted by an infinitesimal factor due
to the contribution of the rest geometric terms, which indeed
behave exactly as a cosmological constant, something which becomes
abundantly clear subsequently. The fact that the initial
conditions for the scalar field were positive imply that the
late-era can indeed be smoothly connected to the inflationary era.
\begin{figure}[h!]
\centering
\includegraphics[width=40pc]{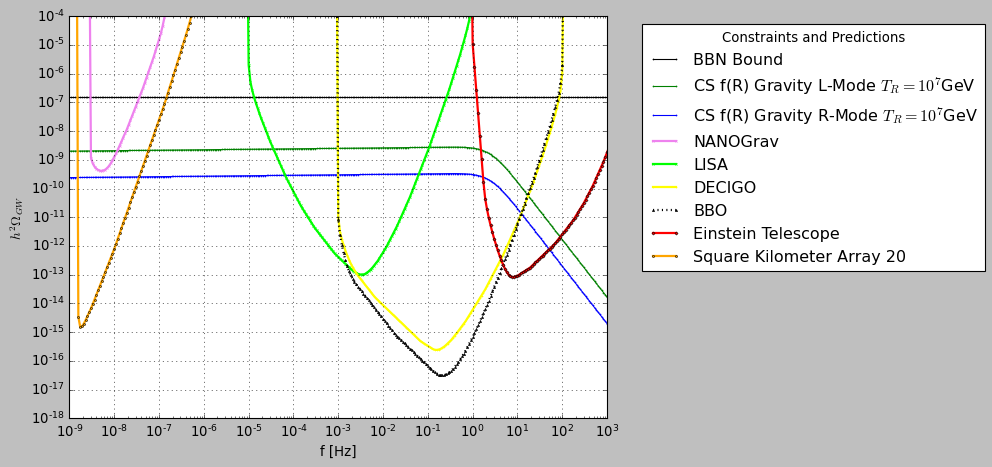}
\caption{Close-up of the $h^2$-scaled gravitational wave energy
spectrum for the Chern-Simons corrected $k$-essence $f(R)$ gravity
for the reheating temperature being $T_R=10^7$GeV.}
\label{plotfinalfrcstr7closeup}
\end{figure}
With regard to the deceleration parameter $q$ in Fig.
\ref{statefinders} we plot its evolution as a function of the
redshift (red curve), and it becomes apparent that the
accelerating era is indeed compatible with latest observations. In
the same plot, we also include the behavior of the deceleration
parameter for the $\Lambda$CDM model. As it can be seen in Fig.
\ref{statefinders}, the model is indistinguishable from the
$\Lambda$CDM model for the whole range of the redshift values
$z=[0,4\times 10^{5}]$. In Table II we quote the values of the
deceleration parameter at present day for the Chern-Simons
$k$-essence $f(R)$ gravity and for the $\Lambda$CDM model. As
shown, the deceleration parameter reaches the value of $q=-1$ in
the future, with the current value also being negative and equal
to $q=-0.52$ and furthermore reaches  the value of $q=1$ for large
values of redshift. Note that a numerical integration in the
interval $[0,4\times 10^5]$ was performed so at order $z\sim10^3$
one observes the deceleration parameter increasing with a
relatively slow rate and going asymptotically to the value of
$q=1$, meaning it transitions from the matter dominated era to the
radiation dominated era smoothly as the redshift increases. Also
in Fig. \ref{omegatotcs} we plot the behavior of the total EoS
parameter as a function of the redshift. As expected it is quite
close to zero during the matter dominated era, as it should, and
starts increasing slowly towards the value of $\omega=\frac{1}{3}$
for $10^3<z<4\times 10^5$, for which interval, the numerical
solution for $y_H$ and $\phi$ was extracted.
\begin{table}[h!]
  \begin{center}
    \caption{\emph{\textbf{Statefinders for the Chern-Simons $k$-essence $f(R)$ model compared to the $\Lambda$CDM Model.}}}
    \label{table2}
    \begin{tabular}{r|r|r}
     \hline
      \textbf{Statefinder Parameter} & \textbf{k-essence $f(R)$  numerical value} & \textbf{ $\Lambda$CDM Value} \\
           \hline
      $q(0)$&-0.51895&-0.535\\ \hline
      $\Omega_{DE}(0)$ & 0.679335 & $0.6847\pm 0.0073$ \\ \hline
      $\omega_{DE}(0)$ & -1 & $-1.018\pm 0.031$\\ \hline
      \hline
    \end{tabular}
  \end{center}
\end{table}
Finally, the dark energy EoS parameter as a function of the
redshift is presented in Fig. \ref{DE}, where it can be seen that
$\omega_{DE}=-1$ for the whole range $z=[0,4\times 10^5]$ and this
is no surprise since the dark energy behaves exactly as a
cosmological constant. Thus the model is validated to mimic
exactly the $\Lambda$CDM model.
\begin{figure}[h!]
\centering
\includegraphics[width=40pc]{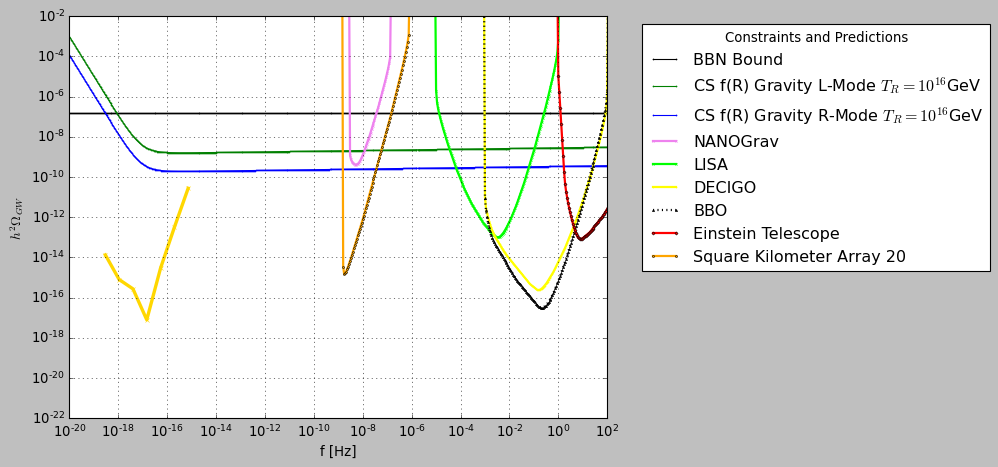}
\caption{The $h^2$-scaled gravitational wave energy spectrum for
the Chern-Simons corrected $k$-essence $f(R)$ gravity for the
reheating temperature being $T_R=10^{16}$GeV.}
\label{plotfinalfrcstr16}
\end{figure}
Furthermore, for completeness in Table II we present the
confrontation of the model with the 2018 Planck constraints
\cite{Akrami:2018odb} on the cosmological parameters or the
comparison of the model with the $\Lambda$CDM model. As it can be
seen, the model is compatible with the Planck data and mimics
closely the $\Lambda$CDM model. Having the numerical solution for
$y_H$ and effectively having the evolution of the Hubble rate
available for the model at hand, makes the calculation of the
parameters (\ref{amchersimonsfinal}) easy and thus the ``damping''
factor $\mathcal{D}$ (\ref{dform}) and in effect
$e^{-2\mathcal{D}}$, can be calculated directly for the
Chern-Simons $k$-essence $f(R)$ gravity model. Having these
available, the calculation of the primordial gravitational wave
energy spectrum can be calculated easily. This is the subject of
the next section.

\subsection{Primordial Gravitational Wave Energy Spectrum}

At this stage, after having solved the Friedmann equation
numerically and having the numerical solution for the Hubble rate
of the Chern-Simons $k$-essence $f(R)$ gravity model at hand, we
can directly proceed in making predictions for the model's
primordial gravitational wave energy spectrum. Again, the spectrum
will be evaluated by using the formula,
\begin{equation}\label{mainfunctionforgravityenergyspectrumcsgravity}
    \Delta_h^2(k)=e^{-2\,\mathcal{D}}r\mathcal{P}_{\zeta}(k_{ref})\left(\frac{k}{k_{ref}}
\right)^{n_T}
    \left ( \frac{\Omega_m}{\Omega_\Lambda} \right )^2
    \left ( \frac{g_*(T_{\rm in})}{g_{*0}} \right )
    \left ( \frac{g_{*s0}}{g_{*s}(T_{\rm in})} \right )^{4/3} \nonumber  \left (\overline{ \frac{3j_1(k\tau_0)}{k\tau_0} } \right )^2
    T_1^2\left ( x_{\rm eq} \right )
    T_2^2\left ( x_R \right ),
\end{equation}
and recall that the primordial inflationary spectrum is evaluated
at the CMB pivot scale $k_{ref}=0.002$$\,$Mpc$^{-1}$, and the
various parameters appearing in Eq.
(\ref{mainfunctionforgravityenergyspectrumcsgravity}) are defined
below Eq. (\ref{mainfunctionforgravityenergyspectrum}). The
``damping'' factor $\mathcal{D}$ is evaluated by the formula
(\ref{dform}) which for the Chern-Simons $k$-essence $f(R)$
gravity takes the form,
\begin{equation}\label{dformcs}
\mathcal{D}=\frac{1}{2}\int^{\tau}a_{M\ell}\mathcal{H}{\rm
d}\tau_1=\frac{1}{2}\int_0^z\frac{a_{M\ell}}{1+z'}{\rm d z'}\, ,
\end{equation}
where recall that for redshifts $z=[0,4\times 10^5]$ the parameter
$a_{M\ell}$ is given by,

\begin{equation}\label{amchersimonsfinalcs}
a_{M\ell}=\frac{f_{RR}\dot{R}+2\lambda_{\ell}\ddot{\nu}k_p/a-2\lambda_{\ell}\dot{\nu}k_pH/a}{(f_R+2\lambda_{\ell}\dot{\nu}k_p/a)}\,
,
\end{equation}
where $k_p$ is the Chern-Simons pivot wavelength to be specified
later on. For redshifts beyond $z\sim 4\times 10^5$, in which case
the cosmological system has entered the radiation domination era,
recall that the parameter $a_{M\ell}$ reads,
\begin{equation}\label{amchersimonsfinalsimpl}
a_{M\ell}=\frac{2\lambda_{\ell}\ddot{\nu}k_p/a-2\lambda_{\ell}\dot{\nu}k_pH/a}{(f_R+2\lambda_{\ell}\dot{\nu}k_p/a)}\,
.
\end{equation}
since $\dot{R}\simeq 0$ for these redshifts. Now, having the
formulas and the numerical solutions available we can proceed and
make predictions for the primordial gravitational wave energy
spectrum for the model at hand. We shall examine in detail how the
primordial gravitational wave energy spectrum is affected by the
$f(R)$, the $G(X)$ and the $\nu(\phi)$ functions, thus how does
the $k$-essence and Chern-Simons terms affect the spectrum of the
inflationary gravity waves today. It turns out that in general
$G(X)$ is subleading compared to function $f(R)$, therefore one
would expect an effective damping in the intensity of
gravitational waves, which is typical for a plethora of $f(R)$
models. However, due to the inclusion of a Chern-Simons term and
in particular the aforementioned toy model of an exponential
scalar coupling function, one can easily evaluate such intensity
for gravitational waves and come to the realization that the
Chern-Simons term manages to enhance such signals up to
$e^{-\mathcal{D}}\simeq 10^{9}$, thus the spectrum is amplified.
Also due to the Chern-Simons term, the two polarization modes do
not propagate equivalently, thus we end up with two distinct
signals for the primordial gravitational waves. Let us analyze in
detail the physical picture, so for the exponential Chern-Simons
coupling of Eq. (\ref{expCS}), and the numerical solution for the
Hubble rate $H(z)$ in Fig. \ref{amLRupdatecs} we plot the behavior
of the parameters $a_{ML}$ (left plot) and of $a_{MR}$ (right
plot) for $z=[0,4\times 10^{5}]$.
\begin{figure}[h!]
\centering
\includegraphics[width=40pc]{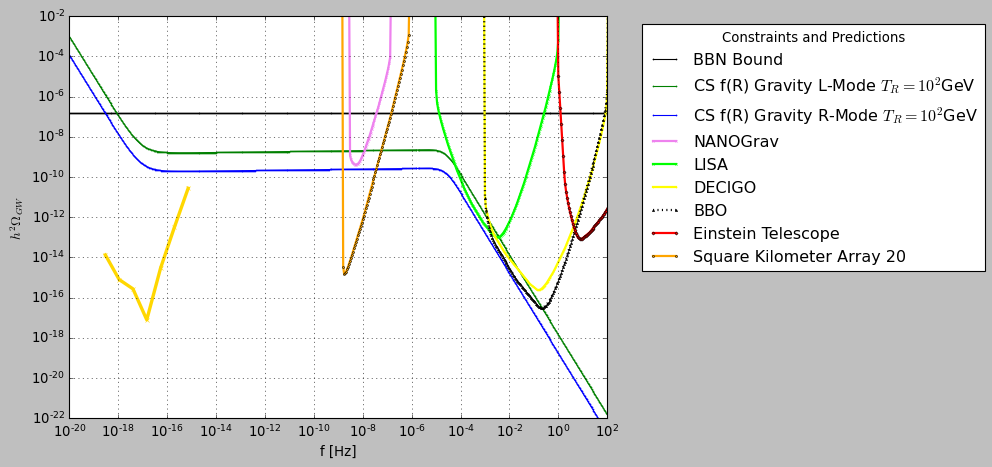}
\caption{The $h^2$-scaled gravitational wave energy spectrum for
the Chern-Simons corrected $k$-essence $f(R)$ gravity for the
reheating temperature being $T_R=10^{2}$GeV.}
\label{plotfinalfrcstr2}
\end{figure}
As it is obvious, firstly the contribution of both the modes will
lead to an overall amplification of the primordial gravitational
wave signal and secondly the two polarizations lead to a different
value of amplification. By computing the ``damping'' factor for
the two polarization modes, we get $e^{-2\mathcal{D}_L}\simeq
7.0267\times 10^{9}$ and $e^{-2\mathcal{D}_R}=8.42676\times
10^{8}$, and note that for the evaluation of the integral we used
a Chern-Simons pivot wavelength of the order $k_p\sim
10^{11}$$\,$Mpc$^{-1}$ which comparable to the wavelengths of the
modes that crossed the horizon for the second time around the
reheating era, and for modes related to the LISA, DECIGO, BBO and
Einstein Telescope. Thus the amplification of the gravitational
wave signal is apparent, and also the two polarizations lead to
different amplifications, thus two signals are expected and this
is indeed what happens. Also our analysis indicated that the
reheating temperature also affects the energy spectrum of the
primordial gravitational waves, as expected from the related
literature \cite{Nakayama:2008wy}. In Figs. \ref{plotfinalfrcstr7}
and \ref{plotfinalfrcstr7closeup} we plot the primordial
gravitational wave energy spectrum for reheating temperature
$T_R=10^{7}$GeV, actually Fig. \ref{plotfinalfrcstr7closeup} is a
close up of Fig. \ref{plotfinalfrcstr7}. Also in Figs.
\ref{plotfinalfrcstr16} and \ref{plotfinalfrcstr2} the primordial
gravitational wave energy spectrum is plotted for the reheating
temperature chosen $T_R=10^{16}$GeV and $T_R=10^{2}$GeV
respectively. We need to note that for all the gravitational wave
spectrum calculations, the tensor spectral index was chosen
$n_T=0.0165879$ and the tensor-to-scalar ratio $r=0.00316538$ and
these values were obtained in the section where the inflationary
era was discussed for the $k$-essence $f(R)$ gravity model with
exponential Chern-Simons coupling.

At this point let us discuss the resulting physical picture, which
seems quite intriguing. Firstly, the gravitational wave spectrum
signal for the Chern-Simons $k$-essence $f(R)$ gravity model at
hand is detectable by almost all the future collaborations, and
for all the reheating temperatures, except when $T_R=10^{2}$GeV in
which case the Einstein Telescope will not capture any signal for
this particular model. Secondly, this specific model, due to the
Chern-Simons coupling and the inequivalent propagation of the two
polarization modes, leads to two distinct signals, which is a
possibility that one could expect in future experiments like the
LISA collaboration etc. Furthermore, the reheating temperature
also affects the spectrum, a feature which was also expected.
Finally, for the model at hand, the BBN bound (\ref{BBNbound}) is
violated only for modes that reentered the horizon quite ``close''
to present time, for frequencies below $10^{-18}$Hz. In fact at
this frequency range, only the LiteBird may detect signals up to
$10^{-18}$Hz, but for the model at hand, as we already mentioned,
below $10^{-18}$Hz the model violates the BBN bound. This effect
however might be due to the fact that we did not take into account
the damping effect caused in the spectrum by neutrino and other
relativistic particles freestreaming below $f\sim 10^{-15}$Hz,
which could lower significantly the gravitational wave energy
spectrum of the model at hand. In any case this is a
toy-pedagogical model that we use in order to extract useful
information for the predictions of modified gravity gravitational
waves energy spectrum for frequencies probed by LISA and other
experiments of the same frequency band, so we do not aim to focus
on low frequencies. The resulting picture by combining the results
of this section and the results of the pure $f(R)$ gravity
section, is quite interesting since simply stated, the plot
thickens with future observations. Simple $f(R)$ gravity models
cannot lead to detectable primordial gravitational waves, however,
exotic terms like the Chern-Simons, may generate a detectable
signal by some or all the collaborations. This feature is mainly
controlled by the tensor spectral index. A positive tensor
spectral index, combined with a reasonable value of the
tensor-to-scalar ratio, will surely be detected by some or all the
future experiments. A negative tensor spectral index will possibly
lead to a non-detectable primordial gravitational wave energy
spectrum signal. Also, one may indirectly determine, or obtain
useful information for, the reheating temperature. This however
might be a model-dependent feature to our opinion. Thus what
should one expect by the detection or the non-detection of the
primordial gravitational waves in future experiments? We list here
some of the most important features:
\begin{itemize}
    \item No signal is detected in future experiments. Bad news for inflation?
    Not necessarily. If the sensitivities of the experiments like
    LISA are higher than the signal, one cannot exclude scalar
    models or modified gravity models like $f(R)$ gravity. One is
    sure in this case, the tensor spectral index is negative, and
    no conclusion can be made for the reheating temperature.
    \item A signal is detected in some or all the future
    experiments. Then one may conclude that a positive tensor spectral
    index is the characteristic of the underlying theory. Since
    the single scalar field theory can yield a blue-tilted tensor
    spectral index only when it is a tachyonic theory, a possible
    detection of a primordial gravitational wave signal will
    exclude single field inflation possibly. The reheating
    temperature is a model dependent feature, so by combining
    results from different experiments, one may conclude if it is
    smaller than, or larger than $T_R=10^{7}$GeV. Now depending on
    the magnitude of the observed spectrum, compared to the GR
    result with a positive tensor spectral index, is it possible
    to conclude whether a modified gravity drives the evolution?
    Possibly no, there are models which we shall present
    elsewhere, for which the spectrum is identical with the GR
    spectrum with blue-tilted tensor spectral index.
\end{itemize}
Thus in conclusion, the plot thickens with primordial
gravitational waves in future observational missions and
experiments. Modified gravity blurs the outcomes to say the least.
The signal is affected mainly from the tensor spectral index and
the reheating temperature only in this case, but no one can say
for sure which modified gravity drives the evolution. Certainly
though if a signal is detected, a non-conventional modified
gravity of some sort must be generating the signal, with a
positive tensor spectral index. If no signal is detected, things
will be as they are now, and the question whether inflation took
ever place cannot be answered unless the sensitivities are
improved in the far future. If in the far future, no signal is
detected again, then possibly inflation must be abandoned as a
candidate for describing the primordial era of our Universe.

\section{Future Perspectives and Concluding Remarks}

In this work we addressed in a quantitative way the primordial
gravitational wave problem in the context of $f(R)$ gravity
related theories. We chose the $f(R)$ gravity models in such a way
so that both inflation and dark energy can be described in a
unified way. Actually for both the models, all the cosmological
eras can be described in a viable way, that is from inflation a
smooth transition is guaranteed to the radiation era, which
gradually evolves to the matter domination era, and the latter
smoothly evolves to the dark energy era. Also both the models we
used, mimic the $\Lambda$CDM model at late times, and one of the
two models is indistinguishable from the $\Lambda$CDM model from
the dark energy back to the radiation domination era. Both the
models studied produce a viable inflationary era which is
compatible with the Planck 2018 constraints on inflation, and both
the models are also compatible with the 2018 cosmological
constraints of the Planck collaboration. We used a WKB formalism
in order to quantify the effect of the modified gravity models and
we arranged the cosmological equations in such a way so that to
facilitate the calculation of the WKB contribution to the
primordial gravitational wave waveform. For the pure $f(R)$
gravity model, the primordial gravitational wave energy spectrum
was slightly enhanced compared to the GR spectrum, however the
predicted spectrum was way below the lowest sensitivities of the
future experiments at large frequencies. This feature has as
source the red-tilted tensor spectral index of the pure $f(R)$
gravity and in order to investigate what would happen in the case
that a blue tensor spectral index was produced, we provided a
Chern-Simons potential-less $k$-essence $f(R)$ gravity model,
which as we showed can produce a blue-tilted tensor spectral
index. After calculating the predicted energy spectrum of the
primordial gravitational waves, we demonstrated that this model
has two interesting features, firstly the presence of two signals
of primordial gravitational waves, and secondly the spectrum is
significantly enhanced compared to GR, thus it can be detected
from all the future experiments on gravitational waves. Another
interesting feature is that the reheating temperature affects the
energy spectrum of the gravitational waves, and actually for
low-reheating temperatures, at high frequencies the energy
spectrum of the gravitational waves drops significantly, and in
some cases, it ceases to be detectable from the future
experiments, like for example from the Einstein telescope.

Also we discussed the possibility of signal detection, or the
absence of a stochastic signal detection in future experiments,
and what would these situations mean for inflation. If no signal
is detected would this mean the end of inflation? Possibly no,
since a red-tilted spectral index would be produced from a scalar
or some other theory, and the signal might lie below the lower
sensitivities of the future experiments. So even more sensitive
future experiments would be required in order to conclude whether
inflation ever took place. So in the case of no signal detection
in all future experiments, the following conclusions can be
reached probably:
\begin{itemize}
    \item Inflation might still be a valid theory for the
    primordial era.
    \item A single scalar field theory, or a pure $f(R)$ gravity
    or any other theory that can produce a red-tilted tensor
    spectral index might actually be responsible for the
    inflationary era.
    \item The tensor spectral index of the primordial tensor
    perturbations is probably red-tilted.
\end{itemize}
If in some future experiments a signal is detected, or if the
signal is detected in all the experiments, then the following
conclusions can be made:
\begin{itemize}
    \item The tensor spectral index is definitely positive.
    \item Single field inflation is excluded, because tachyon
    inflation is bad news for physics.
    \item An exotic modified gravity that can produce a blue-tilted tensor spectral index, certainly drives inflation, perhaps in
    combination with some scalar field theory.
    \item Depending on the detected signal, conclusions on the
    reheating temperature can be made, especially if the signal is
    detected in some future experiments, in a specific frequency
    range, and not in another frequency range.
    \item If the signal is detected in some but not all the future
    experiments, we may seek for further suppression mechanisms in
    the frequencies that no detection occurs. Perhaps another
    mechanism further suppresses the energy spectrum of the
    primordial gravitational waves in a specific frequency range,
    for example supersymmetry breaking. This is a rather exciting
    perspective.
    \item Depending on the magnitude of the signal, one may even
    try to narrow down the modified gravity theories that may
    produce such spectrum. For example, as we will report in some
    time from now, Einstein-Gauss-Bonnet theories do not enhance
    the signal, but produce a blue-tilted tensor spectral index.
    Thus one may theorize whether a stringy theory drives
    inflation or inflation is driven by some more involved
    modified gravity, like the Chern-Simons $k$-essence $f(R)$
    gravity we presented in this paper.
    \item If two signals are detected, then the presence of
    Chern-Simons term is guaranteed to be present during the
    inflationary era.
\end{itemize}
Thus in conclusion, as it was expected, the detection of a signal
in future experiments will put the theoretical minds into fire,
trying to pinpoint the underlying theory that is observed in the
experiments. This perspective is exciting and we surely hope that
this might be the case in the future. Such a result would indicate
that single scalar field inflation might be assisted, not solely
by curvature terms \cite{Oikonomou:2021msx}, but also by more
exotic terms, like the Chern-Simons terms. In the case of
detection of signals, cosmologists are confronted with a demanding
challenge: the models that they use thereafter must provide a
unified description of dark energy and inflation, with a smooth
transition in between them, passing through the radiation and
matter domination eras, a successful late and inflationary
phenomenology, and the model must also produce the right amount of
gravitational radiation in order to comply with the future
detected signals. Exciting and demanding perspective to say the
least. In this work we tried to provide the first models toward
this direction.

To conclude, there are further possibilities for model building
that we did not take into account, like for example the
possibility of having a model with blue-tilted tensor spectral
index that can produce an early dark energy era
\cite{Clarke:2020bil}. This kind of models could also
simultaneously provide insights for alleviating the Hubble tension
problem, but we refrain to go deeply in these problems unless the
calibration issues
\cite{Mortsell:2021nzg,Perivolaropoulos:2021bds} of the Hubble
tension observations are appropriately resolved. In the same
spirit, a possible deviation from a pure radiation domination era
occurring prior to the BBN, can be directly impacted and spotted
on the gravitational wave energy spectrum, for large frequencies,
larger than $f>10^{-11}$Hz. In this case, the phenomenology is
quite rich, even in the context of pure $f(R)$ gravity, a problem
that we will report in due time. A last interesting perspective
that is worthy mentioning is the effect of primordial
non-Gaussianities on gravitational waves, or the induction of
gravitational wave by the non-Gaussianities see for example
\cite{Cai:2018dig}, and  this issue is also worthy studying in a
future work.

\section*{Acknowledgments}

This work is supported by MINECO (Spain), FIS2016-76363-P.

\end{document}